\documentclass[pra,10pt,a4paper,twocolumn,eqsecnum,showpacs]{revtex4}
\usepackage{amsfonts}
\usepackage{amsmath}
\usepackage{amssymb}
\usepackage{bbm}
\usepackage{graphicx}

\newcommand{\bs}{\boldsymbol}
\renewcommand{\bf}{\mathbf}

\begin{document}
\title{Quantum information processing in self-assembled crystals of  cold polar molecules}
\author{M. Ortner$^{1,2}$, Y. L. Zhou$^3$, P. Rabl$^2$, P. Zoller$^{1,2}$}
\affiliation{$^1$Institute for Theoretical Physics, University of Innsbruck, 6020 Innsbruck, Austria}
\affiliation{$^2$Institute for Quantum Optics and Quantum Information of the Austrian Academy
of Sciences, 6020 Innsbruck, Austria}
\affiliation{$^3$College of Science, National University of Defense Technology, Changsha, 410073, China}

%
\date{\today}

\begin{abstract}
We discuss the implementation of quantum gate operations in a self-assembled dipolar crystal of polar molecules. Here qubits are encoded in long-lived spin states of the molecular ground state and stabilized against collisions by repulsive dipole-dipole interactions.  To overcome the single site addressability problem in this high density crystalline phase, we describe a new approach for implementing controlled single and two-qubit operations based on resonantly enhanced spin-spin interactions mediated by a localized phonon mode.  This local mode is created at a specified lattice position with the help of an additional marker molecule such that  individual qubits can be manipulated by using otherwise  global static and microwave fields only. We present a general strategy for generating state and time dependent dipole moments to implement a universal set of gate operations for molecular qubits and we analyze the resulting gate fidelities under realistic conditions.  Our analysis demonstrates the experimental feasibility of this approach for scalable quantum computing or digital quantum simulation schemes with polar molecules. 
\end{abstract}


\maketitle

\section{Introduction}
Rapid progress in control of atomic, molecular and optical (AMO) systems in the recent past has established an experimental basis to realize quantum information processing (QIP) schemes with great success~\cite{revBlatt, revBloch, revRyd, revPM}.
All basic ingredients that are necessary for QIP~\cite{DiVincenzo}, i.e. highly developed state preparation, single particle control and measurement as well as controlled many body gate operations have been experimentally demonstrated with cold atoms and ions, which take a pioneering role in this context.
Prominent examples include ion trap experiments, from early realizations of controlled gate operations~\cite{CnotReal} to recently achieved complex quantum simulations~\cite{IonQSimu}, as well as the addressing of single spins~\cite{locManip} and a variety of many body gate implementations~\cite{collGate, RydbergGate} in systems of neutral atoms.
In recent years, parallel to this development considerable theoretical and experimental efforts have been focused on achieving a similar level of control over polar molecules \cite{coolingEXP2,coolingEXP3,coolingEXP4,coolingEXP5,coolingEXP6,coolingTH1,coolingTH2}.
While state preparation and manipulation of single molecules are still part of ongoing efforts, techniques for the generation of dipolar gases in the quantum degenerate regime have already been demonstrated in the lab \cite{qdegGas1,qdegGas2,qdegGas3,qdegGas4}. The unique features of polar molecules \cite{revPM}, like controllable permanent dipole moments and long lived rotational states with addressability by microwave fields, offer new and interesting possibilities for QIP.

The first quantum computation scheme with polar molecules has been proposed by Dave DeMille~\cite{QIPDeMilleOriginal} with molecules trapped at separate sites of a 1D optical lattice and qubits encoded in the rotational states. In this setup individual site addressability is achieved by spectrally resolving distinct qubits using an electric gradient field and a strong and state dependent dipole-dipole interaction between neighboring molecules is employed to implement two qubit gates. To extend this basic concept beyond an ``always-on" interaction using switchable dipole moments has been proposed~\cite{QIPswitch} with resemblance to QIP concepts with Rydberg atoms in optical lattices \cite{RydbergJaksch}. Other ideas include hybrid quantum computing schemes with atom-molecule platforms~\cite{QIPhybrid1,QIPhybrid2} or molecules coupled to stripline cavities~\cite{QIPhybrid3,AndreNatPhys2006,QIPhybrid4,PRA76}.

In contrast to preceding proposals here we want to focus on QIP in self-assembled molecular dipolar crystals (MDC). Such (quasi-) crystals form under strong repulsive dipole-dipole interactions in an ensemble of polar molecules confined to a plane with perpendicular dipole moments aligned by a strong electric field~\cite{MDC1,MDC2,MDC3}. 
Contrary to Wigner crystals that rely on Coulomb interactions the formation of an MDC is favored by high densities due to the fast decay of the dipole-dipole interaction and lattice spacings range down to $a\sim 100$ nm~\cite{MDC1}. From a perspective of QIP this would be advantageous as it leads to potentially stronger molecular interactions in comparison to previously described setups with optical lattices. In addition, the self-assembled crystal provides a stable defect-free periodic structure even at finite temperatures up to  $\sim\mu$K. Finally, the trapping potential for a MDC only requires a strong transverse confinement and is therefore more easily combined with on-chip magnetic or electrostatic trapping techniques~\cite{AndreNatPhys2006}.
Indeed, based on these properties the storage of ensemble qubits encoded in collective spin excitations of an MDC has previously been proposed~\cite{QIPhybrid3,QIPhybrid4,PRA76,yelinMDC}. However, the high densities  and the much stronger coupling to phonons make the crystalline phase incompatible with conventional schemes for addressing and manipulating qubits encoded in individual molecules of the crystal.

In this work we describe a new approach for QIP which is especially adapted to the physical conditions of an MDC. The main idea is to use the strong phonon coupling to create a controlled phonon mediated interaction for gate operations between qubits instead of the direct dipole-dipole interaction. Similar gate operations that feature such a phonon bus are frequently employed in quantum computing and simulation schemes in chains of trapped ions \cite{PhononIonQC1,PhononIonQC2,PhononIonQC4,PhononIonQC5,CiracPorras}. Here light forces can be used to displace individual ions in a state selective way to create effective interactions between the otherwise highly decoupled internal states. We are interested in similar techniques by creating state dependent dipole moments for molecular qubits using microwave fields only. However, as noted above opposed to the large spacings in ion chains the high densities of MDC's make it difficult to address single sites. To overcome this problem we suggest to combine the proposed technique with the concept of a marker qubit \cite{MarkerIons1,MarkerIons2}, i.e. an additional molecule which is trapped in a separated layer above the crystal. 
Due to strong inter-layer dipole-dipole interactions this marker qubit modifies the local phonon structure of the MDC such that phonon mediated qubit operations can be implemented at pre-specified lattice site in the crystal using otherwise global addressing only.  As the marker moves between different designated sites of the MDC more efficient gate operations between qubits located at distant sites of the crystal are possible in comparison to previous proposals relying on nearest neighbor interactions.

The remainder of the paper is structured as follows. In Sec.~\ref{sec:1} we present a brief overview of MDCs and a qualitative outline of the key ideas of this work. In Sec.~\ref{sec:2} we describe in more detail  the implementation of molecular qubits state with controllable dipole moments and discuss in Sec.~\ref{sec:3} the resulting direct and phonon mediated dipole-dipole interactions. As a specific example  we evaluate in Sec.~\ref{sec:4} the implementation of a local gate operation in a 1D MDC and finally summarize the main results and conclusions of this work in Sec.~\ref{sec:5}.

\section{Motivation and Overview}\label{sec:1}
\begin{figure}
\begin{center}
\includegraphics[width=0.4\textwidth]{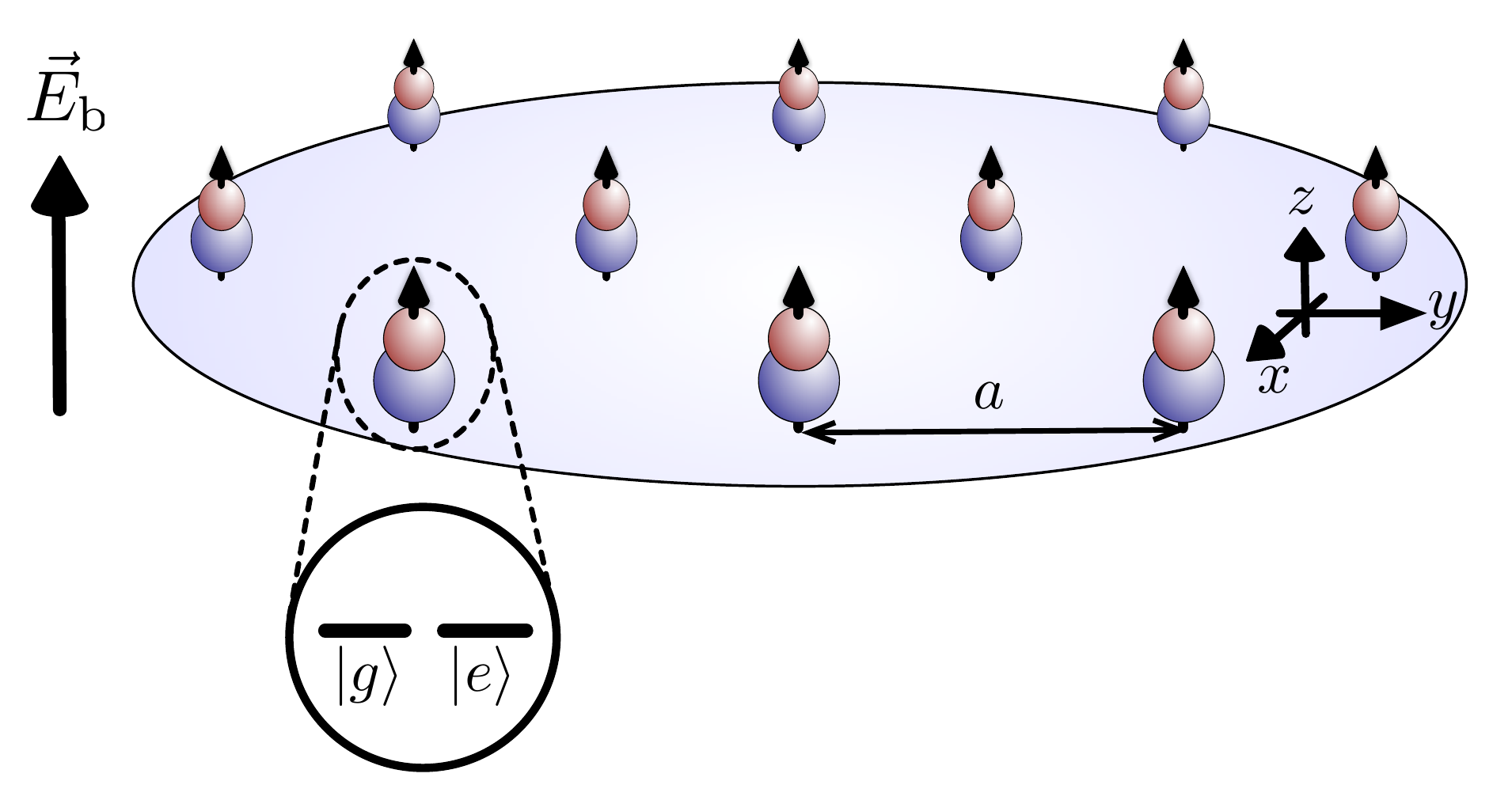}
\caption{A quantum register of molecular qubits confined in a 2D plane and stabilized in a self assembled crystal by strong repulsive dipole-dipole interactions.  The dipole moments of the molecules are aligned by an electric bias field ${\bf E}_b$ and quantum information is encoded in long-lived spin or hyperfine levels of the rovibrational ground state.}
\label{fig:MDC_Qbit2}
\end{center} 
\end{figure}

We describe the implementation of quantum gate operations in a 1D or 2D MDC as shown in Fig.~\ref{fig:MDC_Qbit2},
where qubits are encoded in long-lived spin or hyperfine states of polar molecules and stabilized against collisions by strong repulsive dipole-dipole interactions. We start in  this section with a brief review of the basic properties a MDC followed by a qualitative outline of the main ideas of our proposal.
This overview then serves as a motivation and guideline for the more elaborate calculations detailed in the remaining sections of the paper.

\subsection{Molecular dipolar crystals} 
Compared to neutral atoms the distinguishing feature of polar molecules is a large electric dipole moment of up to several Debye which can be aligned in the presence of an electric bias field.  If in addition the motion of the molecules is confined to the orthogonal plane by a strong transverse trapping potential (see Fig.~\ref{fig:MDC_Qbit2}),  dipole-dipole interaction between molecules are purely repulsive and at low temperatures the system is stabilized against close encounter collisions.
The dynamics of such an ensemble of polar molecules is described 
by the Hamiltonian 
\begin{equation}\label{DCQM:HMDC}
\hat{H}_{\rm MDC}= \sum_i \frac{\bf{p}_i^2}{2m}+ 
\frac{1}{2} \sum_{i \neq j} \frac{D}{|\bf{r}_{i}-\bf{r}_j|^3},
\end{equation}
where $m$ is the mass and  ${\bf r}_i$ and ${\bf p}_i$ the position and momentum operators of the molecules  and $D=\mu^2/(4\pi\epsilon_0)$ for an induced dipole moment $\mu$. 
At low temperatures and for strong transverse confinement the dynamics of the molecules  is  characterized by a dimensionless parameter $r_d$, which is the ratio between interaction and kinetic energy for a mean interparticle spacing $a$,
\begin{equation}\label{DCQM:rd}
r_d=\frac{E_\text{pot}}{E_\text{kin}}=\frac{D m}{\hbar^2 a}.
\end{equation}
At high densities the potential energy dominates and for $r_d\gg1$ the system enters a crystalline phase \cite{MDC1,MDC2,MDC3}, which for  molecular dipole moments of several Debye occurs at typical interparticle spacings $a$ of a few hundred nanometers up to $\sim\mu$m.

In the crystalline regime molecules become localized at fixed equilibrium positions ${\bf r}_0^i$, and, e.g, in 2D form a triangular lattice.
The dynamics of the MDC is then well described by the residual small fluctuations around the respective equilibrium positions, ${\bf r}_i \simeq {\bf r}_i^0+ \hat{\bf u}_i$, which lead to collective phononic excitations of the crystal  \cite{PRA76} and  
\begin{equation}
\hat{H}_{\rm MDC}  \simeq \sum_ k  \hbar \omega_k \hat a_k^\dag \hat a_k.
\end{equation}
Here $\omega_k$ and $\hat a_k$ are the frequencies and bosonic annihilation operators for phonon mode $k$ as we will discuss in more detail in Sec.~\ref{sec:3}.
  

\subsection{Molecular qubits $\&$ spin-spin interactions}

\begin{figure}
\begin{center}
\includegraphics[width=0.45\textwidth]{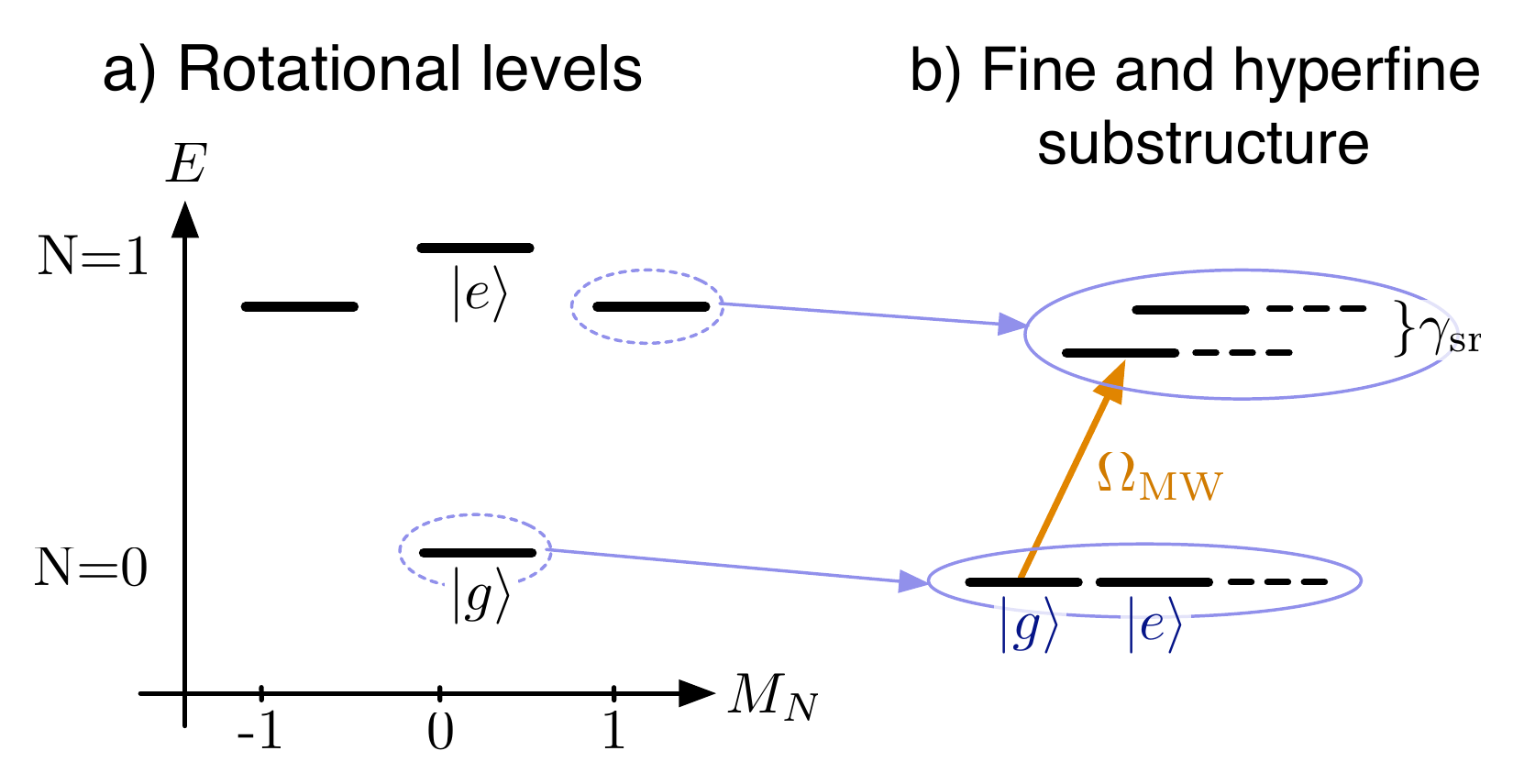}
\caption{ a) Sketch of the rotational spectrum in the presence of an electric bias field and qubit states $|g\rangle$ and $|e\rangle$ are encoded in two long-lived rotational eigenstates of the molecule. b) Qubit states are encoded in two spin or hyperfine sublevels of the rotational ground state. Spin-rotation interactions $\sim \gamma_{sr}$ allow the implementation of state dependent dipole-moments by admixing higher rotational states using microwave fields. See Sec.~\ref{sec:2} for a more detailed discussion.
}
\label{fig:qubits}
\end{center} 
\end{figure}

The MDC represents a stable configuration where in reminiscence of  a Wigner crystal of trapped ions~\cite{revBlatt} the molecules are well localized at individual sites of the self-assembled crystal lattice. 
We are interested in using this crystal as a quantum register where  quantum information is encoded in two long-lived internal states $|g\rangle$ and $|e\rangle$ of the individual molecules. As illustrated in Fig.~\ref{fig:qubits}, rotational spectroscopy \cite{BC} allows us identify different sets of qubit states \cite{PRA76, QIPDeMilleOriginal}, and for example $|g\rangle$ and $|e\rangle$ can be encoded in two rotational eigenstates or in two spin or hyperfine sub-levels of a single rotational manifold.

While in principle both rotational and spin degrees of freedom would provide a suitable choice for long-lived qubit states  we will focus below  explicitly on spin qubits as shown in Fig.~\ref{fig:qubits} b). In this case the idle qubit states have identical dipole moments while appropriate microwave fields can be used to admix higher rotational levels to engineer dipole moments in a state selective way. More precisely we will show in Sec.~\ref{sec:2}  how to generate a state dependent dipole moment which can be written as 
\begin{equation}\label{2:dipoleMoment}
\hat \mu = \mu_0  \mathbbm{1}+ \delta \mu \cdot \hat \sigma_{\vec w}.
\end{equation}
Here $ \hat \sigma_{\vec w} = w_x \hat \sigma_x + w_y \hat \sigma_y + w_z \hat \sigma_z$  is an arbitrary Pauli operator in the qubit space $\{|g\rangle, |e\rangle\}$, specified by the unit vector $\vec{w}$.
To guarantee the stability of the crystal independently of the internal states we require $\epsilon:= \delta \mu/ \mu_0 \ll 1$.  Under these conditions the picture of molecules being localized around their equilibrium positions  ${\bf r}_i \simeq {\bf r}_i^0$ is still valid, but on top of that the state-dependent dipole-dipole coupling leads to additional direct spin-spin interactions of the form 
\begin{align}\label{DCQM:Hspin}
\hat{H}_\text{s-s}=
\frac{\epsilon^2 D}{2}\sum_{i\neq j}Ê  \frac{  \hat \sigma_{\vec w}^i\hat \sigma_{\vec w}^j   }{|{\bf r}^0_{i}-{\bf r}^0_{j}|^{3}}.
\end{align}
In contrast to trapped ions or neural atoms the state dependent dipole-dipole interactions naturally provide a direct qubit-qubit coupling between neighboring molecules. However, while the interaction (\ref{DCQM:Hspin}) contains the basic ingredients for quantum information processing, it is in this form only of limited use for the implementation of quantum gate operations between individual qubits in a crystal. First, (\ref{DCQM:Hspin}) describes global spin-spin interactions induced by homogenous static and microwave fields, and therefore would again require additional local manipulations on the small scale of the MDC lattice spacing $a$.
Second, in writing Eq. (\ref{DCQM:Hspin}) we have ignored fluctuations of the molecules around their equilibrium positions, which induce decoherence, if not appropriately taken into account.

\subsection{Resonantly enhanced phonon mediated dipole-dipole interactions}

To overcome the limitations of direct spin-spin interactions mentioned above, we propose in this work the implementation of enhanced phonon mediated  gate operations which rely on virtual excitations of isolated vibrational modes in the crystal.
To illustrate  the basic idea of this approach we must now take into account  modulations of the dipole-dipole coupling induced by small position fluctuations $\hat {\bf u}_i$ of the molecules. As we show in more detail in Sec.~\ref{sec:3}  these fluctuations result in spin-phonon interactions $\hat{H}_\text{s-p}$ where for $\epsilon \ll 1$  the dominant contribution is of the form 
\begin{equation}
\hat{H}_\text{s-p}\simeq  \sum_{i,k}    \lambda_k^i  \hat  \sigma_{\vec w}^i  (\hat a_k + \hat a_k^\dag).
\end{equation}
Here we have expressed displacements $\hat {\bf u}_i$ in terms of the normal mode operators $\hat a_k$ such that $\lambda_k^i\sim\delta\mu$ is the coupling of phonon mode $k$ to the spin at lattice position $i$.

To make use of the spin-phonon coupling in a controlled manner we introduce a modulation of the dipole moment $\delta \mu(t)=\delta \mu \cos(\omega_0 t)$ with a frequency $\omega_0$. Under certain conditions detailed in Sec.~\ref{sec:3} we can then eliminate the external degrees of freedom and as a result we obtain an additional, phonon mediated spin-spin coupling,  
\begin{equation}\label{eq:HspOverview}
\hat{H}_{\rm s-s}^{\rm pm} \approx \frac{1}{2} \sum_{i\neq j}Ê \left(   \sum_k   \frac{\lambda_k^i \lambda_k^j}{\omega_k-\omega_0} \right) \hat{\sigma}_{\vec w}^i  \hat{\sigma}_{\vec w}^j. 
\end{equation}
We see that by  tuning $\omega_0$ close to a particular phonon frequency $\omega_{k}$ we can in principle amplify the coupling mediated by that specific mode to the extend where it dominates over all other contributions, in particular also over direct spin-spin interactions in Eq.~\eqref{DCQM:Hspin}. In that case the spatial pattern of the resulting spin-spin interaction is solely determined by the mode function of the resonantly enhanced phonon mode. 


\subsection{Marker qubits and local gate operations}
\begin{figure}[t]
\begin{center}
\includegraphics[width=0.4\textwidth]{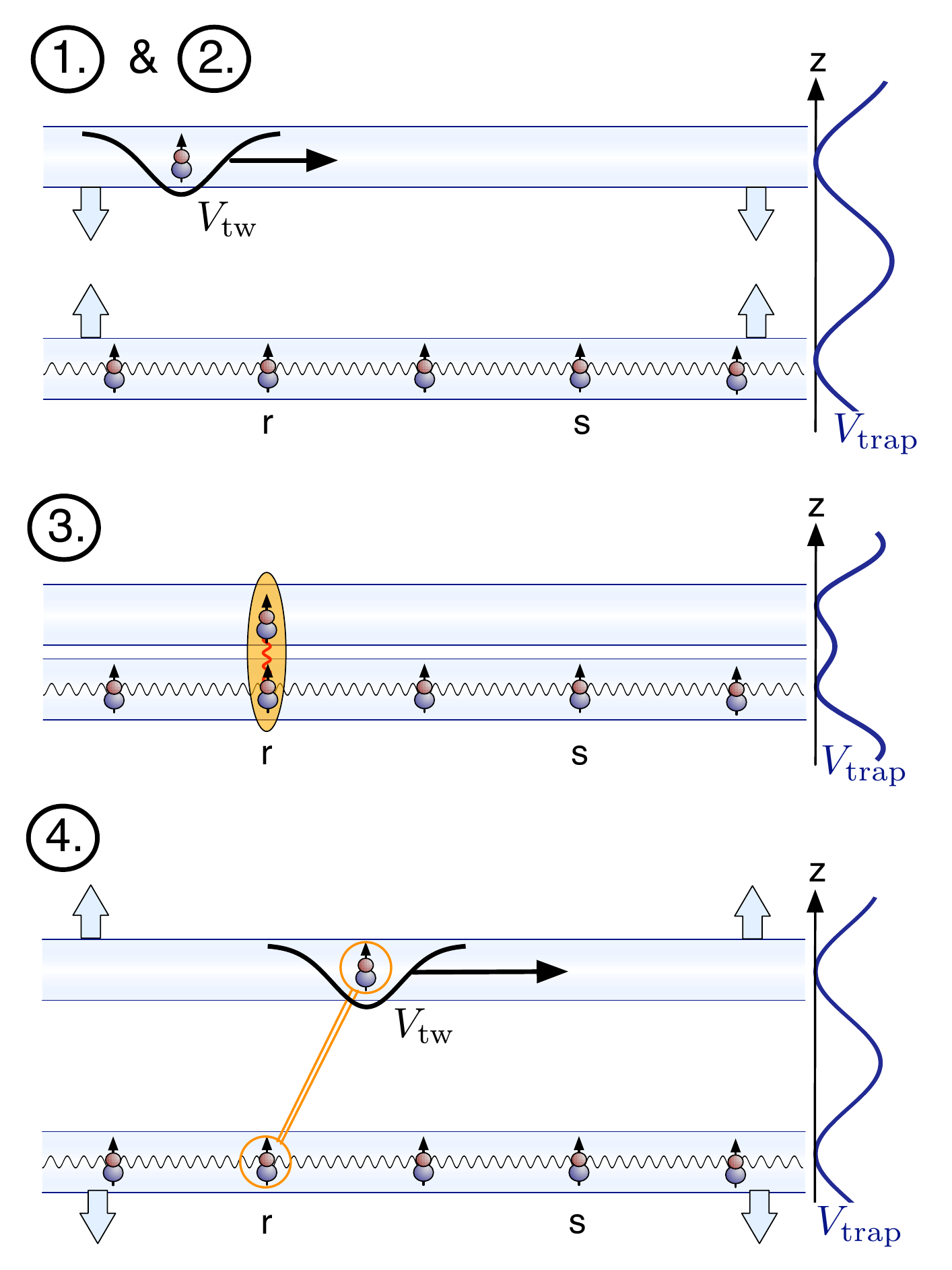}
\caption{Illustration of the basic steps of an entangling operation between the movable marker molecule and a register qubit in the crystal. See text for more details.}
\label{protocol}
\end{center} 
\end{figure}

As a final ingredient for our proposal we now consider an additional ``marker"  qubit, which is independently trapped in a layer parallel to the crystal. By using a weak optical tweezer potential the marker molecule can be moved along the $xy$ plane and locked to a specific register molecule in the MDC via strong attractive interlayer dipole-dipole forces.
This results in the desired formation of a localized phonon mode which ideally involves only a single molecule in the crystal.
Thus, by employing the resonantly enhanced, phonon-mediated spin-spin coupling technique described above, this allows us to implement a general set local two qubit operation between the marker qubit and a target register molecule $r$,  
\begin{equation}\label{1:operation}
\mathcal{U}^r_{G}(\vec w, \phi )= e^{i\phi  \hat \sigma_{\vec w}^m  \hat \sigma_{\vec w}^r}.
\end{equation}
Note that this set of operations is universal and includes single qubit rotations of the register molecule by applying $\mathcal{U}^r_G(\vec w, \phi )$ with the marker qubit prepared in one of the eigenstates of $ \hat \sigma_{\vec w}^m$.

As illustrated in Fig. \ref{protocol} the implementation of arbitrary quantum information processing protocols in a MDC can then be decomposed into the following steps:
\begin{enumerate}
\item The trapping layers for the MDC and the marker molecule are well separated such that the state of the marker qubit can be manipulated  independently of the register qubits in the crystal.  

\item By using a weak optical tweezer potential $V_{\rm tw}(x,y)$ the marker qubit is positioned in the $xy$ plane above a target memory qubit $r$ and the optical barrier in z-direction is reduced adiabatically.   

\item The marker and the register molecule $r$ form a localized phonon mode which is used to implement a local phonon-mediated gate operation $\mathcal{U}^r_G(\vec w, \phi )$ while leaving the rest of the quantum register unaffected. 

\item  The two planes are separated again and the marker molecule can be guided to another lattice site $s$ where the sequence is repeated to generate entanglement between the two memory qubits.

\end{enumerate}
Different generalizations of this basics scheme using multiple marker molecules or multiple layer configuration can be envisioned and would allow for a speed up and a parallelization of QIP protocols. However, the key ingredient in all those schemes is the implementation of local $\hat \sigma_{\vec{w}}\hat \sigma_{\vec{w}}$ operations and in the remainder of this work we now present a more detailed discussion of how this can be achieved in the context of polar molecules.

\section{Molecular qubits and state dependent dipole moments}\label{sec:2}
In this section we start with a discussion of molecular qubit states and show how we can use combinations of static and microwave fields to generate the desired dipole-dipole interactions between two molecules.   
As the internal structure of molecules can vary significantly between different species we will here illustrate the implementation of our scheme for the case of a $^2\Sigma$-molecule, but also outline the general strategy for achieving a similar controllability for other molecules.   

\subsection{Molecular spin qubits}\label{sec:rotStruc}
\begin{figure}
\centering
\includegraphics[width=.5\textwidth]{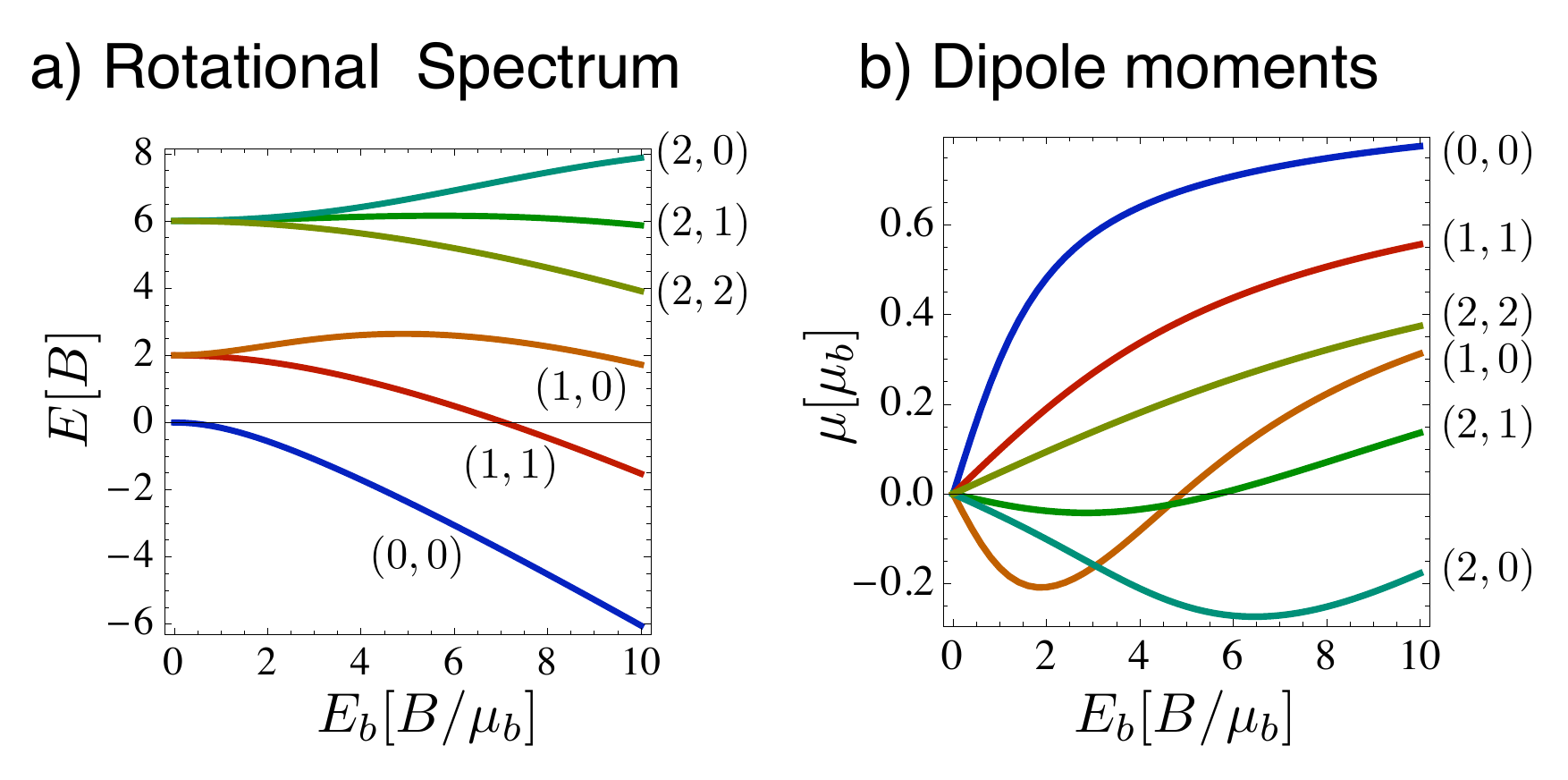}
\caption{Rotational spectrum in the presence of an electric bias field $E_b$. a) Energy levels of $\hat{H}_\text{rot}=B\hat{\bf{N}}^2-\hat \mu_z E_b$ as a function of the bias field $E_b$, where curves for eigenstates $|N,M_N\rangle_{E_b}$ are labeled by $(N,|M_N|)$.  b) The corresponding induced dipole moment $\mu = \, _{E_b}\!\langle N,M_N| \hat \mu_z |N,M_N\rangle_{E_b}$ in units of the bare dipole moment of the molecule $\mu_b$.}
\label{fig:rotSpek}
\end{figure}

With the electronic and vibrational degrees of freedom frozen out, the internal dynamics of cold molecules is well described by its rotational spectrum. For $\Sigma$ ground state molecules this spectrum is usually dominated by the rotational energy, which in the presence of an electric bias field $\bf{E}_b=E_b\bf{e}_z$ is given by $\hat{H}_\text{rot}=B\hat{\bf{N}}^2-\hat \mu_z E_b$. Here $\hat{\bf{N}}$ is the angular momentum of the nuclei, $B$ is the rotational constant which is typically of the order of $\sim 10$ GHz and $\hat \mu_z=\mu_b \hat n_z$ is the z-component of the molecular dipole operator. The bare dipole moment $\mu_b$ depends on the molecular species and can be as large as $8.9$ Debye for SrO. In the limit $E_b\rightarrow0$ the eigenstates $H_\text{rot}$  are the usual  angular momentum eigenstates $|N,M_N\rangle_0$ and we obtain the anharmonic energy spectrum $E_{N,M_N}=B(N+1)N$ of a rigid rotor. At finite, but not too large $E_b$ this level structure is still approximately correct, but the new eigenstates $|N,M_N\rangle_{E_b}$ are now superpositions of states with different $N$ values and therefore exhibit a non-vanishing induced dipole moment $_{E_b}\langle N,M_N| \hat \mu_z |N,M_N\rangle_{E_b}\neq 0$. Fig.~\ref{fig:rotSpek} shows the  energies and induced dipole moments for the lowest rotational states as a function of the applied electric bias field. We see that in principle we can select two of these levels as our qubit state $|g\rangle$ and $|e\rangle$ and by adjusting the bias field induce similar or state dependent dipole moments. However, this would only provide a limited amount of control and for more flexibility we focus in the following on qubits encoded in spin sublevels of a single rotational state. 

For molecules with an unpaired electron the rotational spectrum is in a good approximation described by the Hamiltonian \cite{BC}
\begin{equation}\label{MQSDDM:Hrot}
\hat{H}_{\rm rot}=  B \hat{\bf N}^2   -\hat \mu_z E_b  + \gamma_\text{sr} \hat{\bf N} \hat{\bf S} + \hat{H}_{\rm hyp},
\end{equation} 
where $\hat{\bf S}$ is the electronic spin and $\gamma_\text{sr}$ the spin-rotation coupling constant.  The Hamiltonian $\hat{H}_{\rm hyp}$ accounts for additional hyperfine interactions, which  in the presence of a non-zero nuclear spin $I\neq 0$ lead to a more involved level structure as described below. However,  this additional spin structure  does not affect the main arguments presented here and for simplicity we can assume $I=0$ for the moment. The coupling between spin and rotation $\gamma_\text{sr}\sim 100$ MHz is typically much weaker then the rotational energies and for a sufficiently large bias field $E_b\sim B/\mu$ rotational and spin degrees of freedom are to a good approximation decoupled. 
Therefore, eigenstates of Eq.~\eqref{MQSDDM:Hrot} can be written as $|\psi\rangle \approx |N,M_N\rangle_{E_b}\otimes |S, M_s\rangle$, and, e.g., for $S=1/2$ each rotational level will simply split into a spin doublet. In particular, we now choose two spin sublevels of the rotational ground state for our qubit states $|g\rangle$ and $|e\rangle$ as shown in Fig.~\ref{fig:qubits}.  In this case both states have identical induced dipole moments, which makes these state highly insensitive to noisy electric fields \cite{PRA76}.  Nevertheless, as we will describe in the following, we can use microwave fields to couple $|g\rangle$ and $|e\rangle$ to higher rotational states where a finite admixing of order $\gamma_\text{sr}/B$ between spin and rotation allows us to induce state depend dipole moments.

\subsection{State dependent dipole moments}\label{sec:sddm}
\begin{figure}
\centering
\includegraphics[width=.25\textwidth]{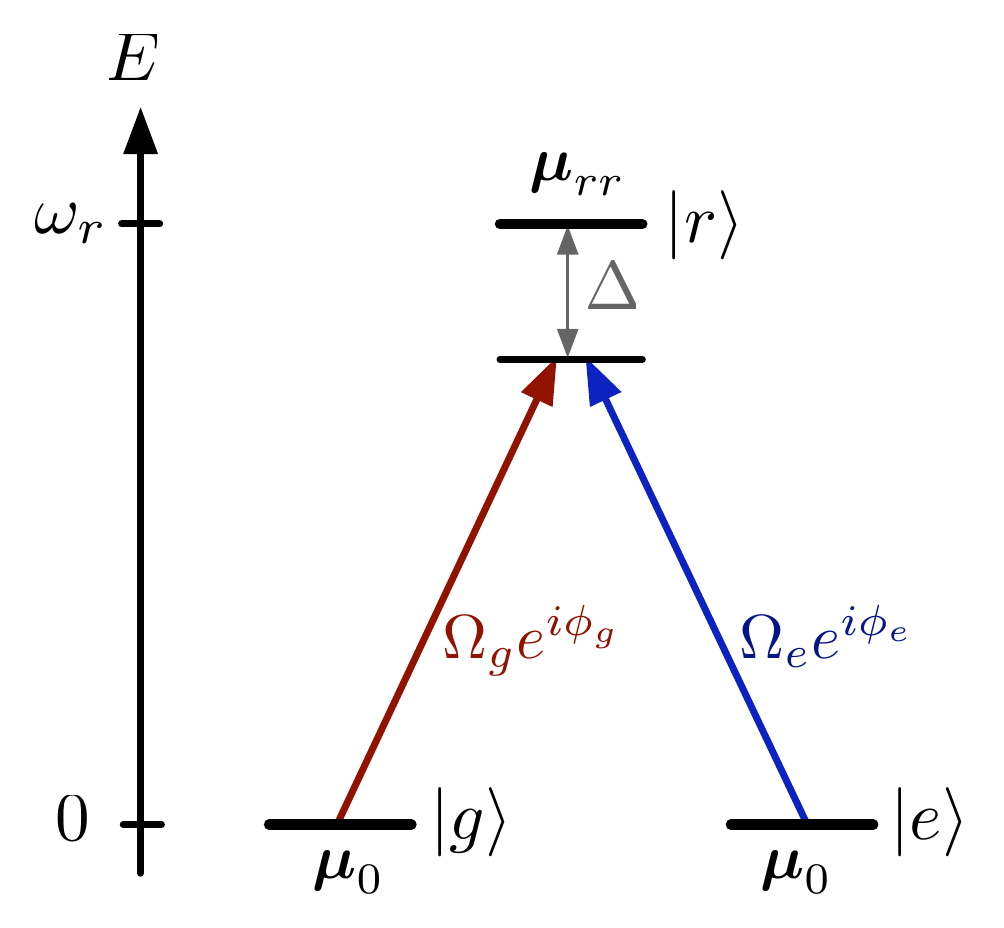}
\caption{General scheme for the creation of a state dependent dipole moment by admixing a rotationally excited level $|r\rangle$ to the qubit states $|g\rangle$ and $|e\rangle$.}
\label{fig:3LS}
\end{figure}

We now outline a general strategy to generate state dependent dipole moments of the form (\ref{2:dipoleMoment}). Restricted to the qubit subspace $\{|g\rangle,|e\rangle\}$ an arbitrary dipole operator is given by
\begin{equation}\label{3:dipoleOperator}
\hat{\bs{\mu}}=\sum_{u,v}\bs{\mu}_{uv}|u\rangle\langle v|,
\end{equation}
where $\bs{\mu}_{uv}=\langle u|\hat{\bs{\mu}}|v\rangle$ and $|u\rangle,|v\rangle\in\{|g\rangle,|e\rangle\}$. By encoding qubit states in  two spin sublevels of the rotational ground state,  we obtain $\bs{\mu}_{gg}=\bs{\mu}_{ee}\equiv\bs{\mu}_0$ and $\bs{\mu}_{eg}=0$ and the internal state dependent part of \eqref{3:dipoleOperator} disappears for such an idle qubit. To introduce non-trivial interactions the states $|g\rangle$ and $|e\rangle$  are coupled by microwave fields \cite{PRA76andrea} to an auxiliary, rotationally excited state $|r\rangle$ as shown in Fig.~\ref{fig:3LS}, similar to \cite{CiracPorras}. This auxiliary level is characterized by an induced dipole moment $\bs{\mu}_{rr}\neq \bs{\mu}_{0}$ and small transition dipole matrix elements $\bs{\mu}_{rg}\equiv\bs{\mu}_{re}\ll \bs{\mu}_0, \bs{\mu}_{rr}$ to eventually suppress unwanted flip-flop interactions as discussed below. We will show in detail how such a three level system can be realized in the following sections.

In the lambda configuration shown in Fig.~\ref{fig:3LS} the states $|g\rangle$ and $|e\rangle$ are coupled to a single rotationally excited state $|r\rangle$ with respective Rabi frequencies $\Omega_g e^{i\phi_g}$ and $\Omega_e e^{i\phi_e}$ and a detuning $\Delta$.
We introduce the ratio of the Rabi frequencies $\tan(\nu)=\Omega_g/\Omega_e$ and $\Omega_\text{eff}=\sqrt{\Omega_g^2+\Omega_e^2}$. The difference between the phases is denoted by $2\eta=\phi_g-\phi_e$. In the frame rotating with the microwave frequency $\omega_\text{MW}=\omega_r-\Delta$, the three eigenstates of this system are given by
\begin{align}
|\alpha_\Omega\rangle &= \cos(\xi_\alpha)|\alpha_0\rangle+\sin(\xi_\alpha)|r\rangle,\\
|\beta_\Omega\rangle &=|\beta_0\rangle,\\
|\gamma_\Omega\rangle &= \cos(\xi_\gamma)|\alpha_0\rangle+\sin(\xi_\gamma)|r\rangle,
\end{align}
where $2\tan(\xi_\alpha)=-(\Delta/\Omega_\text{eff})+\sqrt{4+(\Delta/\Omega_\text{eff})^2}$ and $2\tan(\xi_\gamma)=(\Delta/\Omega_\text{eff})+\sqrt{4+(\Delta/\Omega_\text{eff})^2}$. As $\tan(\xi_\alpha)$ tends to zero when the effective interaction strength disappears, the states $ |\alpha_\Omega\rangle$ and $|\beta_\Omega\rangle$ are adiabatically connected to the superposition states
\begin{align}
|\alpha_0\rangle&=\sin(\nu)e^{i\eta}|g\rangle+\cos(\nu)e^{-i\eta}|e\rangle,\\
|\beta_0\rangle&=\cos(\nu)e^{i\eta}|g\rangle-\sin(\nu)e^{-i\eta}|e\rangle,
\end{align}
which correspond to the so-called bright and dark states of a given set of microwave parameters.

The original qubit states can be expressed in terms of the dressed eigenstates,
\begin{align}
|g\rangle&=e^{-i\eta}\big[\sin(\nu)|\alpha_0\rangle+\cos(\nu)|\beta_0\rangle\big],\\
|e\rangle&=e^{i\eta}\big[\cos(\nu)|\alpha_0\rangle-\sin(\nu)|\beta_0\rangle\big],
\end{align}
which shows how our qubit states are modified when the fields $\Omega_g$ and $\Omega_e$ are adiabatically turned on.
We can now determine the modified dipole matrix elements in \eqref{3:dipoleOperator} in the presence of the fields as
\begin{align}\label{2:dipoleProjector}
\hat{\bs{\mu}}&=\bs{\mu}_0\mathbbm{1}+2\bs{\mu}_1|\alpha_0\rangle\langle\alpha_0|,
\end{align}
where we have defined the dipole moment
\begin{align}
\bs{\mu}_1=\frac{1}{2}\sin^2(\xi_\alpha)(\bs{\mu}_{rr}-\bs{\mu}_0).
\end{align}
By introducing a normalized vector $\vec{w}=(\sin(2\nu)\cos(2\eta),\sin(2\nu)\sin(2\eta),\cos(2\nu))$ the projector $|\alpha_0\rangle\langle\alpha_0|$ can be expressed in terms of the usual Pauli matrices for the qubit states $|g\rangle$ and $|e\rangle$ as
\begin{align}\label{2:}
2|\alpha_0\rangle\langle\alpha_0|=\mathbbm{1}-\hat{\sigma}_{\vec{w}}.
\end{align}
The complete control over our system parameters thus allows us to create an arbitrary state dependent moment $\bs{\mu}_1\hat{\sigma}_{\vec{w}}$ by adiabatically turning on the microwave fields.
The important timescale at which the AC-fields can be varied must be much longer than $1/\Delta$, i.e the time span $\tau$ it takes to turn on the microwave fields to strength $\Omega_\text{eff}$ must fulfill $\tau \gg \Omega_\text{eff}/\Delta^2$.

Finally, we remark that result \eqref{2:dipoleProjector} differs slightly from the purely state dependent moment promised in (\ref{2:dipoleMoment}) as we generate the projector $|\alpha_0\rangle\langle\alpha_0|$ rather than $\hat{\sigma}_{\vec{w}}$. Also we obtain either a positive or  a negative dipole moment $\bs{\mu}_1$, while ideally we would be  interested in generating oscillating dipole moments $\bs{\mu}_1(t)\sim \cos(\omega_0 t)$ with zero mean. These flaws can easily be corrected by using first of all a time depend bias field $\bf{E}_b(t)$ such that the state independent part is absorbed, $\bs{\mu}_0(t)+\bs{\mu}_1(t)=\bs{\mu}_0$. This requires a similar adiabatic condition as above.
Second, to obtain purely oscillating fields we point out that via the relation $\hat{\sigma}_{-\vec w}=-\hat{\sigma}_{\vec w}$ we can change the sign of the state dependent dipole moment by simply changing the microwave field control parameters $\nu$ and $\eta$. Therefore, within the adiabatic regime defined above, the three level configuration shown in Fig. \ref{fig:3LS} is sufficient to engineer 
arbitrary time dependent dipole moments of the form
\begin{align}
\hat{\bs{\mu}}=\bs{\mu}_0\mathbbm{1}+\bs{\mu}_1(t)\hat{\sigma}_{\vec{w}}.
\end{align}
The resulting dipole-dipole interaction between two molecules then becomes
\begin{equation}\label{2:Vdd}
\hat{V}_\text{dd}^{ij}\propto \bs{\mu}_0^2 \mathbbm{1} + \bs{\mu}_0\bs{\mu}_1(t)\left(\hat{\sigma}_{\vec{w}}^i+\hat{\sigma}_{\vec{w}}^j \right)+\bs{\mu}_1(t)^2 \hat{\sigma}_{\vec{w}}^i\hat{\sigma}_{\vec{w}}^j,
\end{equation}
which we will use as a starting point for our discussions of effective spin-spin interactions in Sec. \ref{sec:3} below. 

Note that the form of Eq.~(\ref{2:dipoleProjector}) and in particular the validity of Eq.~(\ref{2:Vdd}) are based on the important condition $\mu_{rg}\equiv\mu_{re}\approx 0$. If this is not the case additional corrections to $\hat{V}_\text{dd}^{ij}$ appear, for example in form of a flip-flop interaction scaling as $\tan^2(\xi_\alpha) |\mu_{er}|^2 \left(  \hat \sigma_+^1\hat \sigma_-^2+\hat \sigma_+^2\hat \sigma_-^1\right)$. To avoid such unwanted contributions we must identify a set of states $|g\rangle$, $|e\rangle$ and $|r\rangle$ where dipole transition matrix elements are  strongly suppressed. In the following we describe in more detail in two examples, how this can be achieved in practice.

\subsection{Example I:  electronic spin qubits}
\begin{figure}
\centering
\includegraphics[width=.45\textwidth]{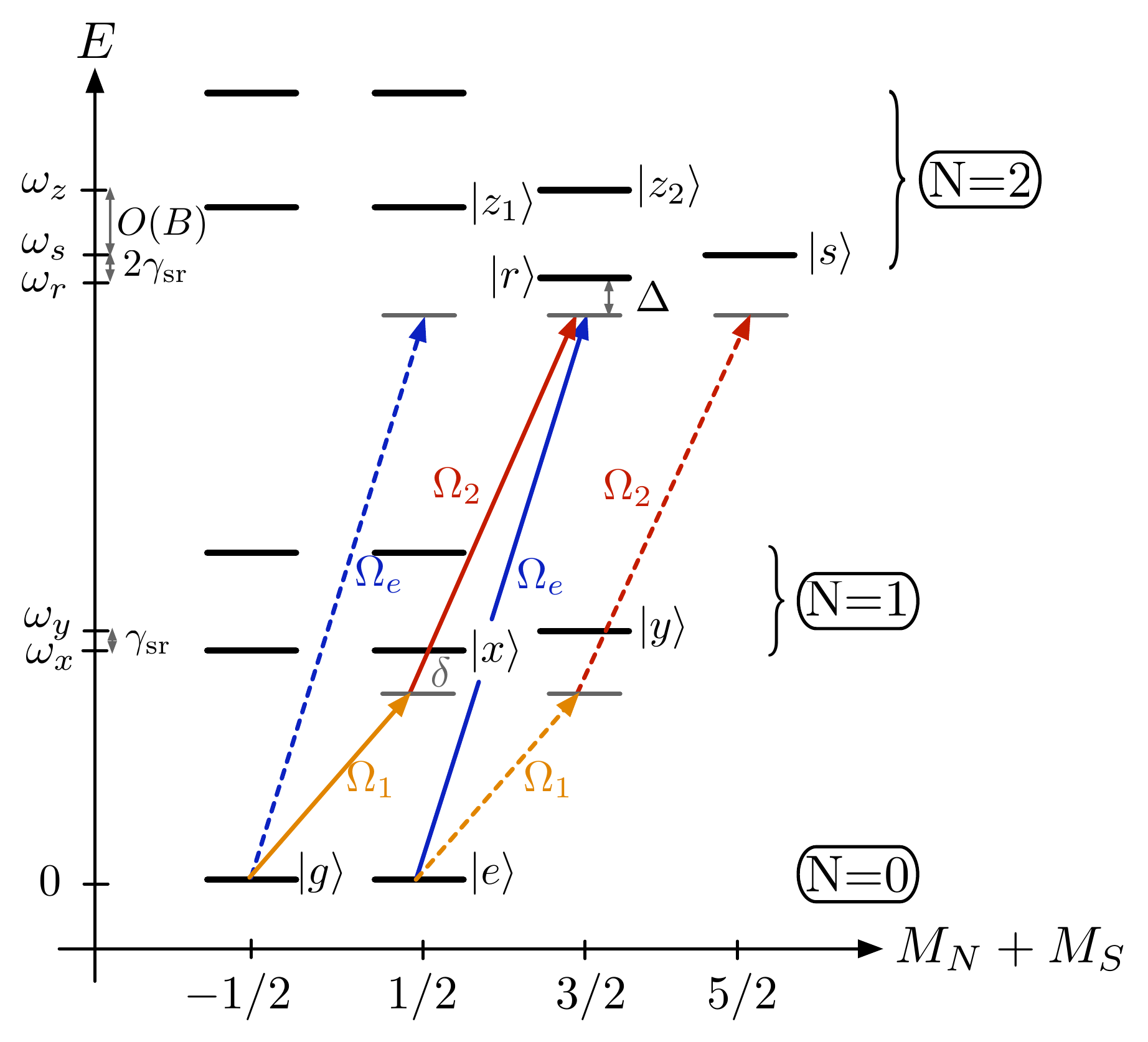}
\caption{Schematic plot of the level structure of ${}^2\Sigma$ molecules. We make use of the specific features of this model resulting from small spin-rotational couplings $\gamma_\text{sr}/B\ll1$ to create an effective $\Lambda$-system between the levels $|g\rangle$, $|e\rangle$ and $|r\rangle$ by applying the microwave fields $\Omega_1$, $\Omega_2$ and $\Omega_e$.}
\label{fig:6LS}
\end{figure}

We first consider the case of a ${}^2\Sigma$ molecules with $S=1/2$ and zero nuclear spin. The lowest states of the resulting rotational spectrum are shown in Fig. \ref{fig:6LS} for a typical value of the spin rotation coupling $\gamma_\text{sr}/B\approx 0.01$. In the following we label eigentstates by $|N,M_N,M_S\rangle$ with the index $E_{b}$ omitted
and we encode our qubit in the two spin states of the rotational ground state,  $|g\rangle=|0,0,-1/2\rangle$ and $|e\rangle=|0,0,1/2\rangle$. In the idle case these two states have identical induced dipole moments, a zero transition dipole moment and are well separated by the rotational splitting $\sim 2B$ from the higher rotational states. 

We identify the state $|2,2,-1/2\rangle$ in the $N=2$ manifold as a suitable level $|r\rangle$ that has zero or spin forbidden transition matrix elements with both $|g\rangle$ and $|e\rangle$, while its induced dipole moment ${\bf \mu}_{rr}$ differs significantly from the one in the ground state.  The state $|e\rangle$ can be coupled directly to $|r\rangle$ with a high ac-field amplitude to overcome the strongly suppressed transition dipole matrix element. For the level $|g\rangle$ we must use a two photon process via the intermediate level $|x\rangle=|1,1,-1/2\rangle$. The proposed microwave fields also couple to additional unwanted levels $|y\rangle=|1,1,1/2\rangle$, $|s\rangle=|2,2,1/2\rangle$ and $|z_{1,2}\rangle$ of our spectrum, as outlined in Fig.~\ref{fig:6LS}.

We are interested in a regime where all contributions from levels other than $|g\rangle$, $|e\rangle$ and $|r\rangle$ are small.
Here the spin-rotation coupling plays a key role. It allows transitions between opposite spin states while at the same time it suppresses unwanted transition matrix elements of spin forbidden transitions by $\gamma_\text{sr}/B$. In addition $\gamma_\text{sr}$ lifts the degeneracy between states $|x\rangle$ and $|y\rangle$ as well as $|r\rangle$ and $|s\rangle$ and enables the design of a state dependent coupling to higher levels.

For $\delta\gg\Omega_1,\Omega_2,\Delta$ we can adiabatically eliminate $|x\rangle$ in the two photon process to give an effective Rabi frequency $\Omega_g=\Omega_1\Omega_2/\delta$ where remaining contributions of the unwanted level are estimated to be of the order of $\Omega_1/\delta$ and $\Omega_2/\delta$ or smaller.
The transition dipole elements between $|e\rangle$, $|y\rangle$ and $|s\rangle$ are similar to the ones between $|g\rangle$, $|x\rangle$ and $|r\rangle$. By making use of the splitting between between the opposite spin states $|x\rangle$ and $|y\rangle$ as well as $|r\rangle$ and $|s\rangle$ the two unwanted levels can be eliminated if the spin rotational coupling fulfills $\gamma_\text{sr} \geq \delta$.
Then the leading contributions from $|y\rangle$ and $|s\rangle$ are of the order of $\Omega_1/\delta$ and $\Omega_1\Omega_2/\delta^2$ respectively.
The ac field that couples $|e\rangle$ directly to the opposite spin state $|r\rangle$ induces additional spin allowed interactions between $|g/e\rangle$ and $|z_{1/2}\rangle$ with large Rabi frequencies of the order of $\Omega_e (B/\gamma_\text{rs})$. However the detuning to these levels, $\omega_z-\omega_s$ is of the order of the rotational constant such that these levels admix only in orders of $\Omega_e/\gamma_\text{rs}$.

Let us now recollect all conditions and parameters involved. 
We require a small spin rotational coupling $\gamma_\text{sr}/B\approx 0.01$ to strongly suppress transition matrix elements between $|e\rangle$ and $|r\rangle$. To eliminate the levels $|y\rangle$ and $|s\rangle$ the detuning $\delta$ must not exceed the spin rotation coupling, $\delta\sim\gamma_\text{sr}$ and the Rabi frequencies must fulfill $\Omega_{1,2}\ll\delta$.
The detuning $\Delta$ should be of the same order as $\Omega_{g,e}$ to obtain reasonable admixings of the upper level. They are then put into relation with other energies by $\Omega_g\approx \Omega_1\Omega_2/\delta$. As the rotational constant is typically of the order of a few to tens of GHz and $\gamma_\text{sr}\approx 100$ MHz we obtain an upper bound for $\Omega_{1,2}$ around $10$ MHz and effective parameters $\Omega_{g,e},\Delta\approx 1$ MHz.
Therefore, the lowest internal energy splitting $\Delta$ in this model still exceeds the relevant dipole-dipole interaction energy $V_\text{dd}\leq 100 \,\text{kHz}$ such that the influence of other molecules can be neglected.


\subsection{Example II:  nuclear spin qubits}
\begin{figure}
\centering
\includegraphics[width=.45\textwidth]{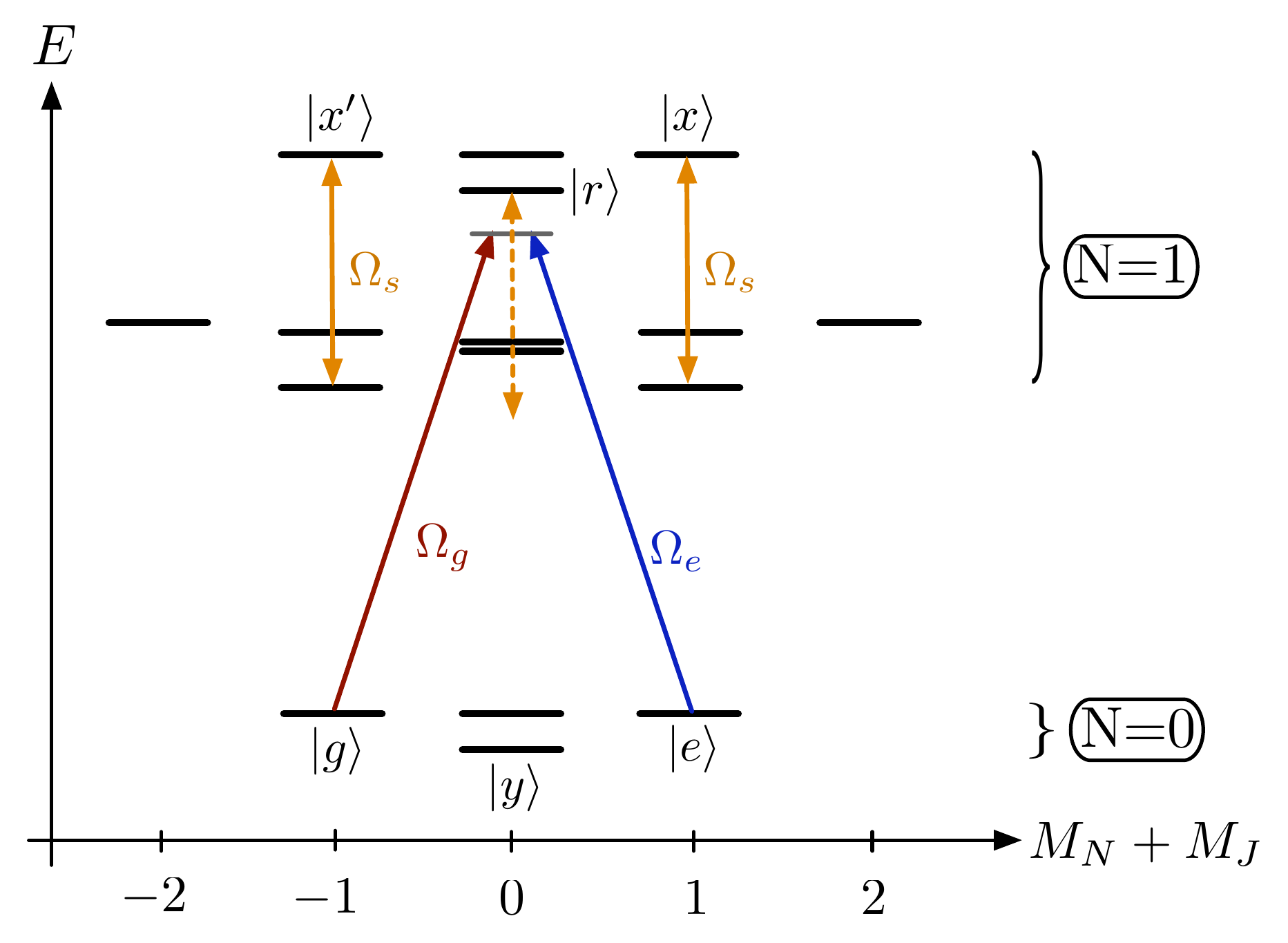}
\caption{Schematic level structure of ${}^2\Sigma$ molecules with a nuclear spin $I=1/2$. The qubit states $|g\rangle$ and $|e\rangle$ are coupled to the auxiliary state $|r\rangle$
via microwave fields of Rabi frequencies $\Omega_g$ and $\Omega_e$. As described in the text, the microwave field $\Omega_s$ is used to introduce additional ac-Stark shifts for the levels $|x\rangle$ and $|x'\rangle$ to suppress otherwise energetically allowed transitions into those states.}
\label{fig:Hyperfine}
\end{figure}

As a second example we now consider molecules with a non-zero nuclear spin, which in Eq.~\eqref{MQSDDM:Hrot} leads to additional hyperfine interactions of the form \cite{BC}
\begin{equation}
\hat{H}_{\rm hyp}= b \hat {\bf S}\hat {\bf I}  + c( \hat{\bf S} \hat {\bf n})( \hat {\bf n} \hat {\bf I}).
\end{equation}
Here $\hat {\bf n}$ is the unit vector along the internuclear axis and $b$ and $c$ are the hyperfine coupling constant which typically are much smaller than $B$, but  comparable to the spin-rotation coupling, $\gamma_\text{sr}\sim |\hat{H}_{\rm hyp}| \sim 2\pi \times 100$ MHz. For the simplest case $S=I=1/2$ we obtain the level diagram shown in Fig.~\ref{fig:Hyperfine}. In particular, we see that the rotational ground state splits into four hyperfine levels corresponding to the eigenstates of the total spin operator $\hat{J}=  \hat{S}+\hat{I}$ and the same structure as above repeats itself in the $|N=1,M_N=0\rangle$ rotational manifold.
Therefore, we can choose the states $|g\rangle=|N=0,M_N=0, J=1,M_J=-1\rangle $ and $|e\rangle=|0,0,1,1\rangle $ as our qubit and the auxiliary, rotationally excited state $|r\rangle=|1,0,0,0\rangle $. In contrast to the spin qubit discussed above, both $|g\rangle$ and $|e\rangle$ are now coupled to $|r\rangle$ via spin-forbidden transitions with strongly suppressed dipole matrix elements $\mu_{er}=\mu_{gr}\approx 0$.

Similarly to the previous model there are various spin allowed transitions  between the $N=0$ an the $N=1$ manifold that lead to resonant flip-flop processes of the form $|r\rangle_1|e\rangle_2 \rightarrow |y\rangle_1|x\rangle_2$ and drive the system outside of our three level subspace. These processes can be energetically suppressed by introducing additional  ac-Stark shifts as indicated in Fig.~\ref{fig:Hyperfine}.

\section{Phonon mediated spin-spin interactions in a dipolar crystal}\label{sec:3}
In the previous section we have shown how microwave fields can be used to manipulate the dipole-dipole interaction between two molecules. We now extend this discussion to molecules embedded in a MDC where state dependent dipolar forces introduce additional interactions between the external and internal degrees of freedom. 
Our main goal is then to show how time dependent control fields can resonantly enhance the coupling to a particular phonon mode and lead to controllable effective qubit-qubit interactions after elimination of the motional degrees of freedom.

The complete Hamiltonian of our model is given by
\begin{equation}
\hat H_{\rm tot}= \sum_i \left(\frac{\bf{p}_i^2}{2m}+\hat H_{\rm rot}^i +V_\text{trap}({\bf r}_i) \right)+ 
\hat{V}_\text{dd}(\{{\bf
r}_{i}\}),
\end{equation}
which includes the kinetic and internal energy molecules, the external trapping potential and the full dipole-dipole interaction
\begin{equation} 
\hat{V}_\text{dd}(\{{\bf
r}_{i}\})=\frac{1}{8\pi\epsilon_{0}}\sum_{i\neq j}\frac{\hat{\bs{\mu}}_{i}\!\cdot\! \hat{\bs{\mu}}_{j}-3({\bf
n}_{ij}\!\cdot\!\hat{\bs{\mu}}_{i})({\bf n}_{ij}\!\cdot\! \hat{\boldsymbol{\mu}}_{j})}{|{\bf r}_{i}-{\bf r}_{j}|^{3}}\,,
\end{equation}
with $\bf{n}_{ij}=\bf{r}_{ij}/|\bf{r}_{ij}|$ and $\bf{r}_{ij}={\bf r}_{i}-{\bf r}_{j}$.

Under conditions detailed in the previous section we can restrict our discussion to dipole operators of the form 
\begin{equation}
\hat{\bs{\mu}}= {\bf e}_z\left(  \mu_0 \mathbbm{1} + \delta  \mu\ \! \hat \sigma_{\vec w} \right),
\end{equation}
with all dipole moments aligned along the $z$ axis.
It is then convenient to split the dipole-dipole interaction into two parts as $\hat{V}_\text{dd}(\{{\bf r}_{i}\})=\hat{V}_\text{dd}^E(\{{\bf r}_{i}\}) + \hat{V}^I_\text{dd}(\{{\bf r}_{i}\})$, where $\hat{V}_\text{dd}^E(\{{\bf r}_{i}\})\propto \mu_0^2$ depends only on the molecular positions while $\hat{V}^I_\text{dd}(\{{\bf r}_{i}\})\propto \delta\mu,\delta \mu^2$ contains the internal state dependent part of the interaction.
We are interested in the regime where a stable crystalline phase is realized independently of the internal configuration which is the case when the stabilizing force $V_\text{dd}^E(\{{\bf r}_{i}\})$ dominates over $V_\text{dd}^I(\{{\bf r}_{i}\})$.
Then the molecular position operators are given by ${\bf r}_i= {\bf r}_i^0+\hat{\bf u}_i$ where $ \hat{\bf u}_i$ denote small fluctuations around the equilibrium positions ${\bf r}_i^0$ and the total Hamiltonian can be decomposed into three parts,
\begin{equation}\label{3:HtotSplit}
\hat{H}_{\rm tot}= \hat{H}_{\rm MDC} +\hat{H}_\text{s} + \hat{H}_\text{s-p}.
\end{equation}
Here the first term represents the external molecular motion and describes the formation of the crystal and its vibrational excitations. This term has already been introduced in Eq. (\ref{DCQM:HMDC}) for a purely 2D configuration and is in general given by 
\begin{equation}
\hat{H}_{\rm MDC}= \sum_i \frac{\bf{p}_i^2}{2m}+V_\text{trap}({\bf r}_i)+ 
\frac{D}{2} \sum_{i \neq j} \frac{1-3\cos(\theta_{ij})}{|\bf{r}_{i}-\bf{r}_j|^3},
\end{equation}
where $\theta_{ij}$ is the angle between $ \bf{r}_{i}-\bf{r}_j$ and the $z$ axis.
The second term contains internal degrees of freedom only and is given by
\begin{equation}\label{PMI:Hspin}
\hat{H}_\text{s}= \sum_i \hat{H}_{\rm rot}^i  + \hat{V}^I_\text{dd}(\{{\bf r}^0_{i}\}).
\end{equation}
Note that a residual spin dynamics from $\hat{H}_{\rm rot}^i$ can be absorbed by applying additional global $B$-fields and we can neglect this term in the following discussion. Finally, the last term in Eq.~(\ref{3:HtotSplit}) contains all remaining contributions to the dipole-dipole interaction which affect both internal and external degrees of freedom. It is defined as $\hat{H}_\text{s-p}=\hat{V}^I_\text{dd}(\{{\bf r}_{i}\})-\hat{V}^I_\text{dd}(\{{\bf r}^0_{i}\})$ and leads to non-trivial spin-phonon interactions, which will be the main focus of our discussion below. 

\subsection{Phonons}
The basis of our description is the formation of a molecular crystal in our model described by the crystal Hamiltonian
\begin{equation}
\begin{split}\label{3:HMDCphonons}
\hat{H}_{\rm MDC} = &\sum_i \frac{\bf{p}_i^2}{2m}+V_\text{trap}({\bf r}_i) + \frac{1}{2} \sum_{i,j} v_\text{dd}({\bf r}_i-{\bf r}_j),
\end{split}
\end{equation}
with $v_\text{dd}({\bf r})= D(1-3 n_{z}^2)/r^3 $ and $D=\mu_0^2/4\pi\epsilon_0$.
In the crystalline phase we can follow the standard procedure \cite{Mahan} and expand the potential to second order in the small displacements $\hat{\bf u}_i$ around equilibrium positions $\bf{r}_i^0$ that are determined by minimizing the potential energy in Eq. (\ref{3:HMDCphonons}). While in homogeneous 1D or 2D systems the $\bf{r}_i^0$ would  form a regular lattice in the general case details depend on the external trapping potentials as described in more detail in Sec.~\ref{sec:4}.
The expansion of Hamiltonian (\ref{3:HMDCphonons}) is given by
\begin{equation}\label{3:HMDC}
\hat{H}_{\rm MDC} \approx \sum_i \frac{\bf{p}_i^2}{2m}+ \frac{1}{2}\sum_{i,j}   \hat{u}_{i,\alpha} \Phi^{\alpha,\beta}_{ij} \hat{u}_{j,\beta},
\end{equation}
where $\Phi^{\alpha,\beta}_{ij}=\partial_{r^\alpha_i}\partial_{r^\beta_j}[V_{\rm dd} (\{{\bf r}^0_{i}\})+V_{\rm trap} (\{{\bf r}^0_{i}\})]$. To diagonalize the quadratic Hamiltonian we  introduce a set of bosonic mode operators $\hat a_k^\dag$ that are related to the displacements $\hat{u}_{i,\alpha}$ by
\begin{equation}
\hat{u}_{i,\alpha}=\sum_k\zeta_k^{i,\alpha}\sqrt{\frac{\hbar}{2m\omega_{k}}}\left(\hat{a}_{k}^\dag+\hat{a}_{k}\right).
\end{equation}
The normalized modefunctions $\zeta_k^{i,\alpha}$ and eigenfrequencies $\omega_k$ are obtained from the resulting eigenvalue problem and we can write the crystal Hamiltonian in terms of the bosonic operators as
\begin{equation}
\hat{H}_\text{MDC}\approx\sum_{k}\hbar\omega_{k} \hat{a}_{k}^\dag\hat{a}_{k}.
\end{equation}

\subsection{Spin-spin and spin-phonon interactions}
In presence of the control fields the dipole-dipole interaction $\hat V_\text{dd}^I(\{\bf{r}_i\})$ gives additional molecular couplings that strongly depend on the internal degrees of freedom.
To maintain the stable crystalline phase that has been described previously such additional interactions must be small compared to the stabilizing potential $ V_\text{dd}^E(\{\bf{r}_i\})$ such that the molecular displacements from the original equilibrium positions fulfill $|\langle \hat{\bf{u}}_i\rangle|\ll a$ for arbitrary internal configurations.
This is realized when
\begin{align}
\frac{\delta\mu}{\mu_0}\equiv \epsilon\ll1.
\end{align}
The state dependent part of the dipole-dipole interaction is given by
\begin{align}
\hat{V}_\text{dd}^I(\{\bf{r}_i\})=\epsilon\sum_{i\neq j} v_\text{dd}(\bf{r}_{ij})\hat{\sigma}_{\vec{w}}^i+\frac{\epsilon^2}{2}\sum_{i\neq j} v_\text{dd}(\bf{r}_{ij})\hat{\sigma}_{\vec{w}}^i\hat{\sigma}_{\vec{w}}^j
\end{align}
and can be expanded to leading order in $\hat{\bf{u}}_i$. 
The lowest order of the expansion gives the spin Hamiltonian (\ref{PMI:Hspin})  
\begin{align}
\hat{H}_\text{s}=\sum_i \frac{B_\text{eff}}{2}  \hat \sigma_{\vec w}^i + \frac{\epsilon^2 D}{2}\sum_{i\neq j}Ê  \frac{  \hat \sigma_{\vec w}^i\hat \sigma_{\vec w}^j   }{|{\bf r}^0_{i}-{\bf r}^0_{j}|^{3}},
\end{align}
where we have introduced an effective magnetic field $B_\text{eff}=  \sum_{j\neq i} 2Ê\epsilon D /|{\bf r}^0_{i}-{\bf r}^0_{j}|^{3}$. 
The remaining part of the expansion leads to spin-phonon interactions, and to first oder in $\hat{\bf{u}}_i$ we find
\begin{align}\label{PMI:Hsp}
\hat{H}_\text{s-p}&\simeq \epsilon \sum_{k,i}\lambda_k^i\hat{\sigma}_{\vec{w}}^i(\hat{a}_k^\dag+\hat{a}_k)+\frac{\epsilon^2}{2}\sum_{k,i\neq j}\kappa_k^{ij}\hat{\sigma}_{\vec{w}}^i\hat{\sigma}_{\vec{w}}^j(\hat{a}_k^\dag+\hat{a}_k).
\end{align}
Here we have introduced
\begin{align}
\kappa_k^{ij}=\sqrt{\frac{\hbar}{2m\omega_{k}}} \vec{v}_{\text{dd}}'(\bf{r}_{ij}^0)\big(\vec{\zeta}_{k}^{i}-\vec{\zeta}_{k}^{j}\big),
\end{align}
where $\lambda_k^i=\sum_{j(\neq i)}\kappa_k^{ij}$ 
and the derivative of the dipole-dipole potential is 
\begin{align}
\vec{v}_\text{dd}'(\bf{r})=\frac{3D}{r^4}\left(\begin{array}{c}   n_{x} (5n_{z}^2-1)\\n_{y}(5n_{z}^2-1) \\n_{z}(5n_{z}^2-3) \end{array}\right).
\end{align}

\subsection{Effective spin-spin interactions using oscillating dipole moments}
Up to this point $\hat{H}_\text{s-p}$ describes an uncontrolled coupling of the spins to the crystal phonons.
It is our aim to amplify the interaction with modes around a chosen frequency $\omega_0$. 
To achieve this we let the small dipole moment $\delta\mu$ oscillate as
\begin{equation}
\delta\mu(t)=\delta\mu\cos(\omega_0 t),
\end{equation} 
with the picture of a classical harmonic oscillator that is driven close to its resonance frequency in mind.
This can be achieved by time-varying control parameters under conditions detailed in Sec. \ref{sec:sddm}.
We proceed in a rotating frame with respect to $\hat{H}_{\omega_0}=\sum_{k}\hbar\omega_0 \hat{a}_{k}^\dag \hat{a}_{k}$, where the total Hamiltonian (\ref{3:HtotSplit}) takes the form
\begin{align}
\hat{\tilde{H}}_\text{tot}=&\sum_k\hbar\Delta_k\hat{a}_k^\dag \hat{a}_k+\epsilon\cos(\omega_0 t)\sum_{i\neq j} v_\text{dd}(\bf{r}_{ij}^0)\hat{\sigma}_{\vec{w}}^i\nonumber\\
&+\frac{\epsilon^2}{2}\frac{1+\cos(2\omega_0 t)}{2}\sum_{i\neq j} v_\text{dd}(\bf{r}_{ij}^0)\hat{\sigma}_{\vec{w}}^i\hat{\sigma}_{\vec{w}}^j\nonumber\\
&+\frac{\epsilon}{2}\sum_{k,i}\lambda_k^{i}\hat{\sigma}_{\vec{w}}^i \big(\hat{a}_k^\dag+\hat{a}_k^\dag e^{i2\omega_0 t}+h.c.\big)+O(\epsilon^2 \hat{\bf{u}}_i).\nonumber
\end{align}
Here the $\Delta_{k}=\omega_{k}-\omega_0$ denote the detunings between $\omega_0$ and the phonon frequencies $\omega_k$.
Note that we have neglected terms of the order $O(\epsilon^2 \hat{\bf{u}}_i)$ that arise from the second term on the righthand side of Eq. (\ref{PMI:Hsp}). Eventually such contributions only give corrections of the order of $O(\epsilon^4)$, and -- as they only couple to low frequency modes -- they are not enhanced by the resonance conditions discussed in the following. By making a rotating wave approximation we obtain
\begin{align}\label{PMI:Hrwa}
\hat{\tilde{H}}_\text{tot}&\simeq  \sum_{k}\hbar\Delta_{k} \hat{a}_{k}^\dag\hat{a}_{k} 
+\frac{\epsilon^2}{4}\sum_{i\neq j} v_\text{dd}(\bf{r}_{ij}^0)\hat{\sigma}_{\vec{w}}^i\hat{\sigma}_{\vec{w}}^j\nonumber\\
&\hspace{2cm}+\frac{\epsilon}{2}\sum_{k,i}\lambda_k^{i}\hat{\sigma}_{\vec{w}}^i \big(\hat{a}_k^\dag+\hat{a}_k\big),
\end{align}
which is valid for
\begin{equation}
v_\text{dd}(\bf{r}_{ij}^0),\ \kappa_k^{ij}\ \ll\ \frac{\omega_0}{\epsilon}.
\end{equation}
This essentially limits our choice of $\omega_0$ to frequencies that are of the order of the dipole-dipole interaction $D/a^3$ or above.

Based on the effective Hamiltonian (\ref{PMI:Hrwa}) we now eliminate the phonons by a canonical transformation.
This Lang-Firsov or polaron transformation \cite{LFT, Mahan} is defined by $\hat{H}\rightarrow\hat{U} \hat{H}\hat{U}^\dag$ where 
\begin{equation}
\hat{U}=e^{\hat{S}}\ \ \ \ \text{with}\ \ \hat{S}= \sum_{i,k}\frac{\epsilon\lambda_k^i}{2\hbar\Delta_k}\hat{\sigma}_{\vec{w}}^i(\hat{a}_k-\hat{a}_k^\dag).
\end{equation}
This transformation changes into a picture where internal and external degrees of freedom are decoupled as $\hat{H}_\text{tot}=\hat H_{\rm MDC}+\hat H_{s,{\rm eff}}+E_\text{p}$, where
\begin{equation}\label{3:HtotFin}
\hat H_{s,{\rm eff}} = \frac{\epsilon^2}{4}\sum_{i\neq j} v_\text{dd}(\bf{r}_{ij}^0)\hat{\sigma}_{\vec{w}}^i\hat{\sigma}_{\vec{w}}^j
+ \frac{\epsilon^2}{2}\sum_{i\neq j,k}\frac{\lambda_k^i\lambda_k^j}{\hbar\Delta_k}\hat{\sigma}_{\vec{w}}^i\hat{\sigma}_{\vec{w}}^j.
\end{equation}
 The phonon coupling has thus been eliminated in favor of an additional spin-spin interaction, the phonon mediated interaction (PMI) $\hat V_\text{pm}$, and a constant energyshift $E_\text{p}=\sum_{i,k}{\lambda_k^i}^2\epsilon^2/2\hbar\Delta_k$ known as the polaron shift. 
The first term in (\ref{3:HtotFin}) is what remains of the direct spin-spin interaction.

\subsection{Discussion}\label{sec:discussion}
\begin{figure}
\centering
\includegraphics[width=.35\textwidth]{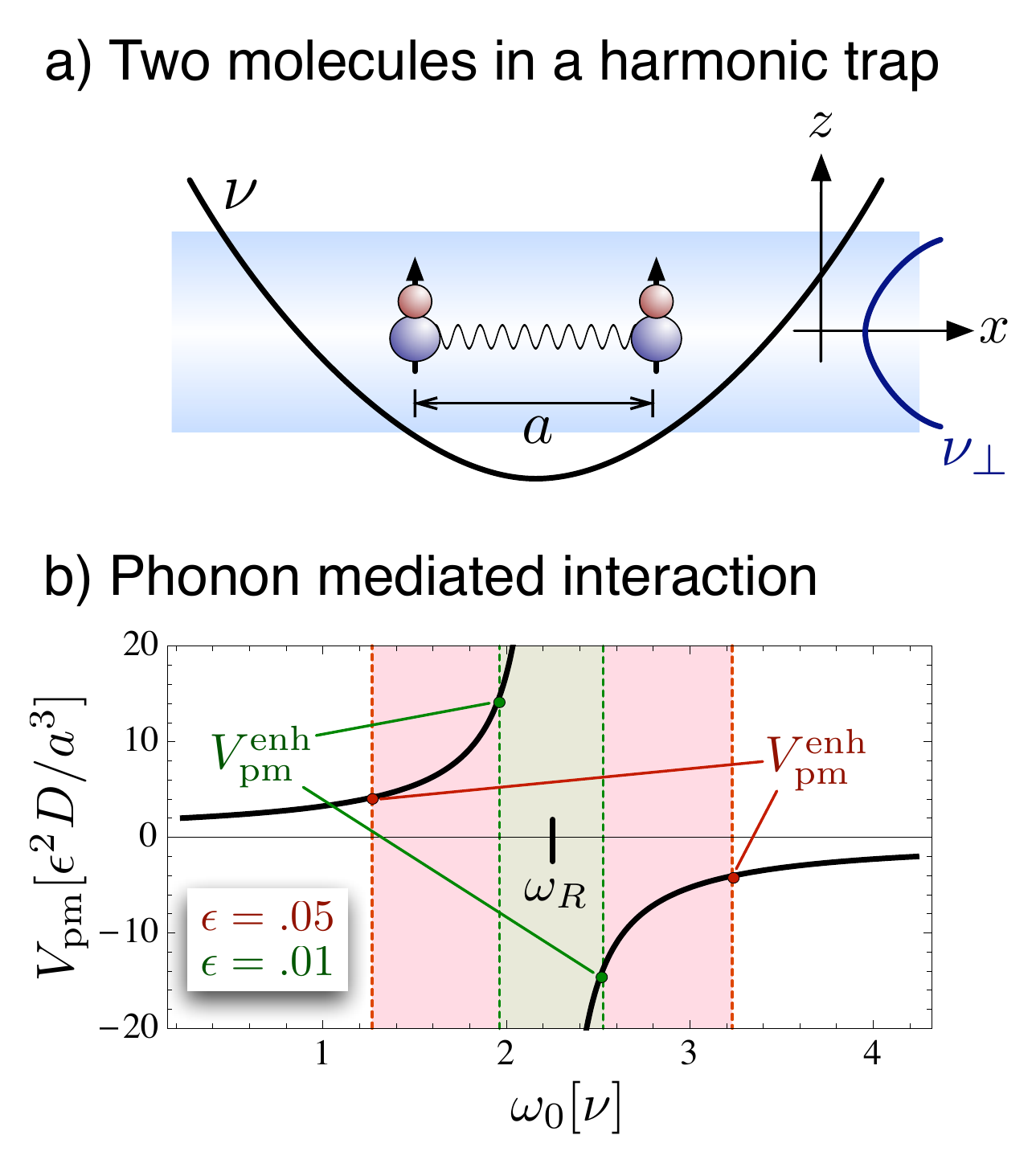}
\caption{a) Phonon mediated interactions between two polar molecules confined to the $x$-axis by strong perpendicular confinement $\nu_\perp$ and with dipole moments aligned along the $z$ axis. The system is stabilized in $x$-direction by a harmonic trapping potential $\nu\ll\nu_\perp$.
 b) The strength $V_\text{pm}$ of the phonon mediated interaction in this system with $\omega_0$ tuned close to the frequency of the longitudinal breathing mode $\omega_R=\sqrt{5}\nu$. In red (green) we outline the forbidden parameter regime for $\omega_0$ when the molecular displacements are restricted to  $\Delta\bar{u}(\epsilon) \ll 1$ for different dipole ratios $\epsilon$. Note that the direct spin-spin interaction in this model has the value of $1/2$ in units of $\epsilon^2 D/a^3$.}
\label{fig:PMIinTARP}
\end{figure}

The phonon mediated interaction can be enhanced by reducing the detuning $\Delta_k$ for a certain phonon mode. This is of course only valid up to a certain point, as we can see from the additional collective molecular displacements 
\begin{equation}\label{5:UaU}
\hat{U}\hat{a}_k^\pm\hat{U}^\dag=\hat{a}_k^\pm- \sum_{i}\frac{\epsilon\lambda_k^i}{2\hbar\Delta_k}\hat{\sigma}_{\vec{w}}^i.
\end{equation}
We introduce an upper bound to the relative displacement for any internal configuration as $\Delta u=\max \{|\hat{\bf{u}}_{i}-\hat{\bf{u}}_{i+1}|\}$.
Then the phonon mediated interaction can be enhanced as long as $\Delta u$ does not exceed a certain fraction of the mean interparticle spacing,
\begin{equation}\label{3:smallCollDisp}
\Delta u=\Big| \sum_k (\vec{\zeta}_k^{i}-\vec{\zeta}_k^{i+1})\sqrt{\frac{\hbar}{2m\omega_k}}
\sum_i\Big|\frac{\epsilon\lambda_k^i}{\hbar\Delta_k}\Big|\Big| \ll a.
\end{equation}
This condition connects $\epsilon$ and $\Delta_k$ such that smaller dipole ratios $\epsilon$ allow us to choose $\omega_0$ closer to resonance with a particular phonon mode of frequency $\omega_{R}$, as illustrated in Fig. \ref{fig:PMIinTARP}b) for a toy model specified below.
For a spectrum that is sufficiently sparse around the resonant mode,  we can achieve  $\Delta_R\ll\Delta_k,\ \forall\ k\neq R$ such that the phonon mediated interaction is to a good approximation given by
\begin{equation}\label{3:VpmApp}
\hat{V}_\text{pm}=\frac{\epsilon^2}{2}\sum_{i\neq j} \frac{\lambda_R^i\lambda_R^j}{\hbar\Delta_R}\hat{\sigma}_{\vec{w}}^i\hat{\sigma}_{\vec{w}}^j
+O\Big[\frac{\Delta_{R}}{\Delta_{k}}\Big].
\end{equation}
As the sign of $\Delta_R$ can be freely chosen by tuning $\omega_0$ above or below the respective resonance frequency $\omega_R$ the PMI can be made attractive or repulsive.
By selecting a fixed value for $\Delta u$ we connect $\epsilon$ and $\Delta_R$ such that the PMI becomes effectively of first order in $\epsilon$.
In this manner the ratio of the direct interaction over the phonon mediated one can be made arbitrarily small by variation of the dipole ratio which of course comes  at the cost of reducing the absolute value of the interaction strength.

Other properties of the PMI are revealed when studying $\lambda_{k}^i\propto\sum_j \vec{v}_\text{dd}'(\bf{r}_{ij}^0)(\vec{\zeta}_k^{i}-\vec{\zeta}_{k}^{j})$ which is nonzero only if there is relative motion between molecule $i$ and any other molecule $j$ in the direction of a nonzero gradient of $v_\text{dd}(\bf{r}_{ij}^0)$ when oscillating in mode $k$. Therefore, dominant contributions to the phonon mediated interaction arise from  the high (low) energy longitudinal (perpendicular) modes in MDC's. In addition, every term in the sum  is suppressed by a factor $1/(r_{ij}^0)^4$, which generally leads to a fast decay of PMI. 

In the following we briefly describe a simple system of two molecules in a harmonic trap as shown in Fig.~\ref{fig:PMIinTARP}a), which illustrates the basics features of our model.
The molecules with dipole moments aligned along the $z$-axis are confined to the $x$-axis by a strong transversal trapping and $V_\text{trap}^i= m\nu_\perp^2(y_i^2+z_i^2)/2+m\nu^2x_i/2$, with $\nu_\perp> \nu$.
The equilibrium spacing along the $x$-axis  is given by $a=(6\mu_0^2/m\nu^2)^{1/5}$.
In this setup the oscillations in $x$, $y$ and $z$ directions decouple with two modes in every direction. They include a center of mass (COM) mode at the respective trapping frequencies and in each direction a breathing mode with respective frequencies $\omega_x/\nu=\sqrt{5}$, $\omega_y/\nu_\perp=\sqrt{1-\nu/\nu_\perp}$ and $\omega_z/\nu_\perp=\sqrt{1-3\nu/\nu_\perp}$.
The eigenvectors in every direction are simply $(1,1)/\sqrt{2}$ for the COM mode and $(1,-1)/\sqrt{2}$ for the breathing mode.

The COM modes do not contribute to the PMI as there is no relative motion between the molecules.
Also the perpendicular breathing modes do not contribute because in our setup the field gradient in $z$ and $y$ directions disappears along the $x$-axis.
The only mode that mediates the PMI is the longitudinal breathing mode and the resulting PMI strength is plotted in Fig.~\ref{fig:PMIinTARP}b). By optimizing the detuning for a fixed value of $\Delta u/a= \Delta\bar{u}$ we obtain
\begin{equation}\label{3:VpmTrap}
\hat{V}_\text{pm}^\text{enh}= \pm \epsilon\Delta\bar{u}\frac{3D}{2a^3}\hat{\sigma}_{\vec{w}}^1\hat{\sigma}_{\vec{w}}^2.
\end{equation}
As pointed out above this enhanced interaction depends linearly on the ratio $\epsilon$ in contrast to the $\epsilon^2$ scaling of direct spin-spin interactions.

\section{Marker molecules and local gate operations in a 1D crystal}\label{sec:4}
\begin{figure}
\centering
\includegraphics[width=.45\textwidth]{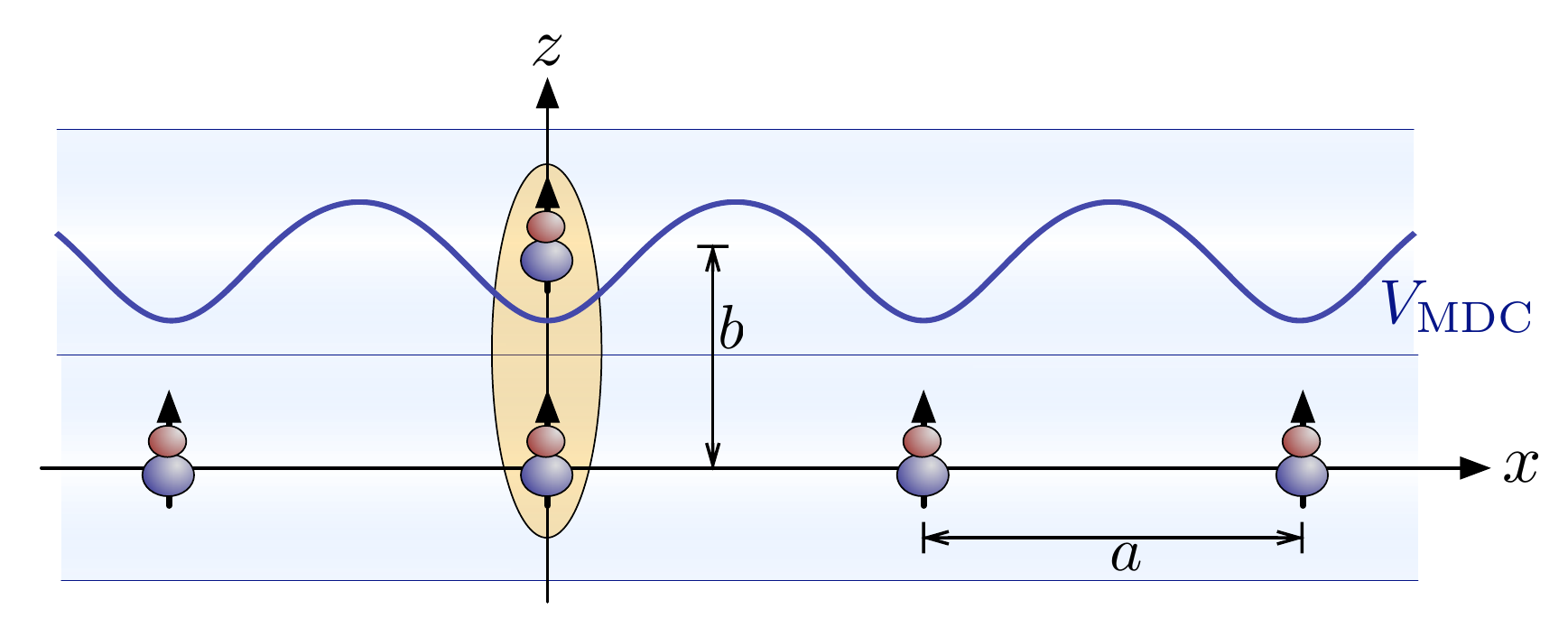}
\caption{Schematic setup for a double layer configuration where an optical standing wave potential confines the MDC and a single marker molecule in two planes separated in $z$-direction by $b=\lambda/2$. The MDC has a lattice spacing $a$ and for dipole moments aligned along ${\bf e}_z$ generates an attractive periodic potential $V_\text{MDC}$ for the marker molecule. For $b/a\lesssim 1$ the marker molecule locks onto the target register molecule and modifies the local phonon structure. 
}
\label{fig:pinSetupNew}
\end{figure}

In the previous section we have described how we can resonantly enhance phonon-mediated dipole-dipole interactions by tuning the oscillation frequency $\omega_0$ close to the frequency of a particular mode phonon mode. We now want to discuss in more detail how this technique can actually be exploited to generate local gate operations between qubits in a large dipolar crystal. To do so we focus on the setting shown in Fig. \ref{fig:pinSetupNew}.
In this double layer configuration
a single ``marker" molecule located in the upper trapping layer can be moved in the $xy$ plane by a weak optical tweezer potential $V_{\rm tw}(x,y)$. By bringing both layers close together, the marker molecule will lock to one of the molecules in the crystal and thereby create a spatially localized phonon mode. We can use this mode to implement a local gate between the marker and the target register molecule. Note that while below we will for simplicity only discuss a 1D model, all arguments and conclusions presented in this section are equally valid for the 2D case. 

\subsection{Local phonon modes of the bilayer system}
We first consider the case where the two optical trapping layers are brought close together and the tweezer potential is switched off.
The minimal layer separation $b$ is then fixed by the wavelength of the trapping laser, $b=\lambda/2$.
Within the first layer polar molecules with dipole moments aligned perpendicular to the trapping plane experience \emph{repulsive} dipole-dipole interactions and form an MDC with average interparticle distance $a$. 
The marker molecule resides in the second layer and sees a periodic potential $V_\text{MDC}$ generated by the crystalline structure. Since this interaction is \emph{attractive} the marker molecule `locks' onto one of the molecules in the lattice below. This configuration has been analyzed in detail in Ref. \cite{MyNJP} where it has been shown that the tunneling rate of such particles to neighboring sites is exponentially suppressed for $r_d\gg 1$ and $b/a<1$~\cite{MyNJP}.


The presence of the additional marker molecule distorts the lattice and modifies the phonon spectrum. However, already simple estimates show that down to values of $b/a\approx 0.5$  the influence of the marker molecules on the lattice structure as well as the crystal phonon spectrum is relatively small due to the anisotropy and the fast decay of the dipole-dipole coupling.
Therefore, as a main effect we expect the emergence of localized phonon modes due to the strong coupling between the marker and a single register molecule.

\begin{figure}
\centering
\includegraphics[width=.45\textwidth]{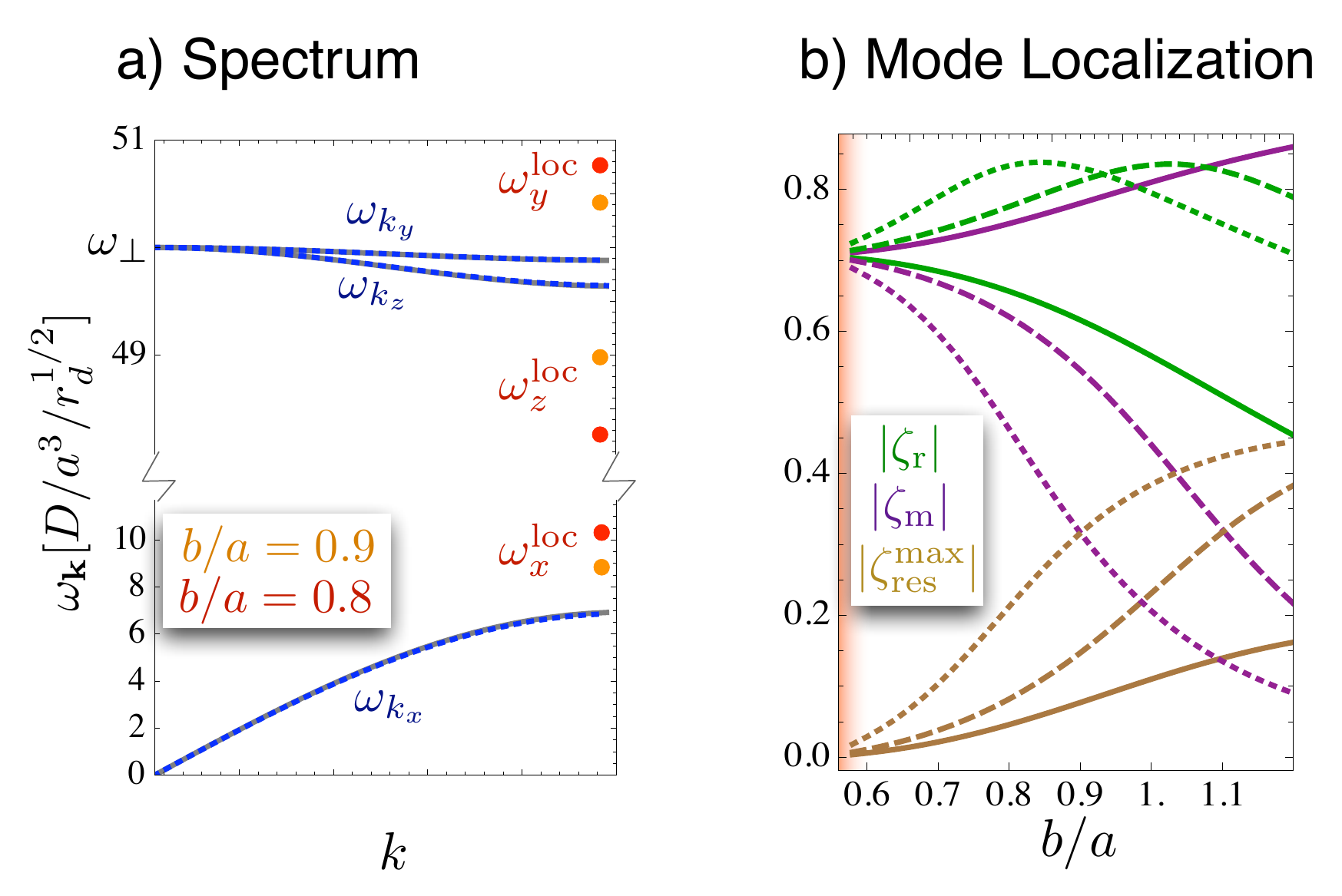}
\caption{
a) Phonon spectrum for a homogenous 1D MDC in presence of a marker molecule as shown in Fig.~\ref{fig:pinSetupNew} for a perpendicular trapping frequency $\omega_\perp=50 D/a^3/r_d^{1/2}$.
Oscillations in different directions decouple and result in two optical and one acoustical branch. This is shown in the presence of the marker (blue, dashed) for $b/a=0.8$ and without the marker (gray, solid).
The three local modes $\omega_{x}^\text{loc}$, $\omega_{y}^\text{loc}$ and $\omega_{z}^\text{loc}$ are displayed in orange (red) for $b/a=0.9\, (0.8)$. 
b) The localization of $\omega_{x,y,z}^\text{loc}$ is visualized by plotting mode functions of the marker (purple), the target molecule (green) and the maximum of all others (brown). The dotted, solid and dashed lines represent values for the different modes $\omega_x^\text{loc}$, $\omega_y^\text{loc}$ and $\omega_z^\text{loc}$ respectively and the red area outlines the limit of the perpendicular stability for this setup \cite{MyNJP}.}
\label{fig:localization}
\end{figure}

As an example we plot in Fig. \ref{fig:localization} a) the phonon spectrum of a 1D dipolar crystal in the presence of the marker molecule.
For this plot we have evaluated the modified equilibrium positions for this trapping configuration and solved numerically for the eigenvalues $\omega_k$ for a system of $N=50$ particles assuming periodic boundary conditions.
The acoustic phonon branch corresponds to  longitudinal vibrations of the lattice and the two optical phonon branches describe transversal excitations with frequencies that are mainly determined by the transverse confinement potential, i.e., $\omega_{k_{y,z}}\approx \omega_\perp$.
We further observe the appearance of three modes $\omega^{\rm loc}_{x,y,z}$ with frequencies that are separated from the rest of the spectrum and correspond to isolated phonon modes which are localized around the marker molecule.
This is illustrated more clearly in Fig.~\ref{fig:localization}b) where we plot the components of the modefunction $\zeta_k^i$ for the modes $\omega^{\rm loc}_{x,y,z}$. As we bring the layers closer together these modes become more and more localized and eventually reduce to relative oscillations between the marker and the target register molecule only.

\subsection{Implementing local gate operations}\label{sec:totalIntinourModel}
The ability to generate a spatially localized and energetically separated phonon mode allows us to use the phonon mediated coupling technique described in Sec.~\ref{sec:discussion} to implement a controlled gate operation between the marker and the target register molecule. 
To do so the modulation frequency $\omega_0$ is tuned close to one of the local mode frequencies, e.g $\omega_z^\text{loc}$
to enhance the local PMI over the global direct spin-spin coupling.
Since in general both interactions will be present, to discuss the fidelity of the local gate operations we write the effective spin Hamiltonian (\ref{3:HtotFin}) as
\begin{equation}
\hat{H}_{\rm s,eff}= U_0 \hat{\sigma}_{\vec w}^\text{m}\hat{\sigma}_{\vec w}^\text{r} + \frac{1}{2} \sum_{i\neq j} {'} \,U_{ij} \hat{\sigma}_{\vec w}^i\hat{\sigma}_{\vec w}^j, 
\end{equation}
where the sum runs over all molecule pairs, excluding the the index pairs $(m,r)$. Our goal is to identify the conditions under which $U_0 \gg U_{\rm res}$ where $U_{\rm res}:={\rm max} \{ |U_{ij}|\}$ is the strength of unwanted residual spin-spin interactions.

We now focus on the setting described above where the marker molecule locks onto the register molecule such that $\bf{n}_\text{mr}=\bf{e}_z$. Then the only mode that contributes to the local interaction is $\omega_z^\text{loc}$ as oscillations in different directions decouple in this configuration similarly to the example given in Sec. \ref{sec:discussion}. This mode has the additional advantage that up to a first approximation it does not contribute to the interaction between other molecules in the crystal.
\begin{figure}
\centering
\includegraphics[width=.5\textwidth]{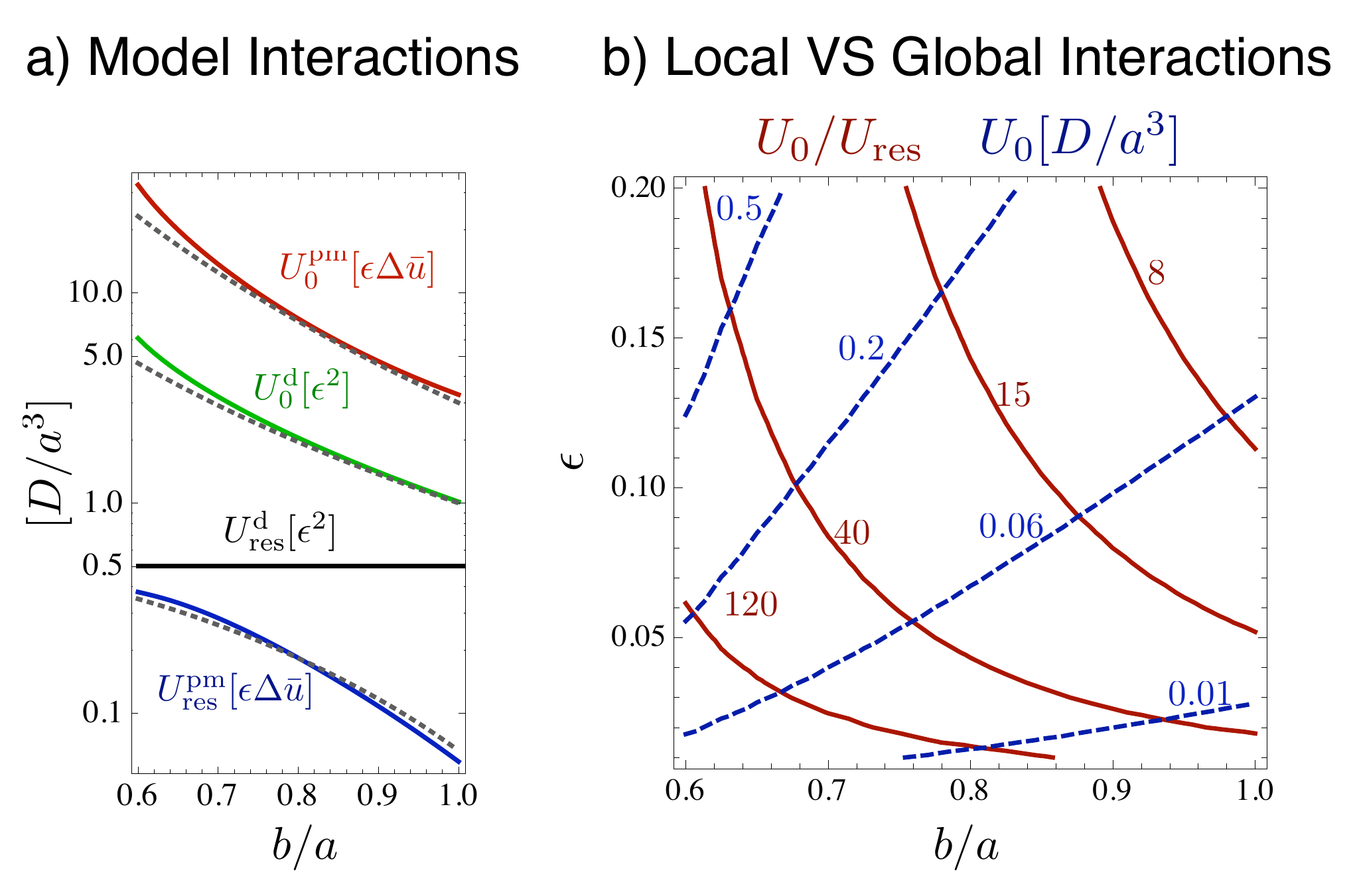}
\caption{a) Different contributions of direct and enhanced phonon mediated interactions to $U_0$ and $U_\text{res}$ for a similar system as outlined in Fig. \ref{fig:localization}. The PMI is enhanced about $\omega_z^\text{loc}$ and gray dotted lines show the respective analytic approximations.
A contour plot of the ratio $U_0/U_\text{res}$ is given in b) for the specific choice of maximal displacement $\Delta u/a=0.1$. For comparison the absolute value of $U_0$ is plotted in the same figure to illustrate the regime of interest.}
\label{fig:pmi3}
\end{figure}

We can derive analytic approximations for the resulting interactions under the assumptions that i) only marker and pinned molecule oscillate with all other molecules at rest and ii) that the marker does not significantly perturb the crystal lattice with molecular equilibrium positions exactly at the center of the transverse trap.
This allows us to write the total local interaction approximately as
 \begin{equation}\label{4:U0}
U_0\approx \frac{D}{a^3} \left(\frac{\epsilon^2}{(b/a)^3} \pm 3 \frac{\epsilon\Delta \bar{u} }{(b/a)^4} \right).
\end{equation}
Here the two contributions arise form the direct coupling and the enhanced phonon-mediated interaction respectively for a specific choice of $\Delta \bar{u}=\Delta u/a$ respectively. Under the similar assumptions we can show that the magnitude of the residual global spin-spin interactions is given by
 \begin{equation}
U_{\rm res} \approx \frac{D}{a^3} \left(\frac{\epsilon^2}{2}\pm\epsilon\Delta \bar{u}\frac{3}{4}\frac{2(b/a)^3-3(b/a)}{(1+(b/a)^2)^{7/2}} \right),
\end{equation}
which is dominated by $(D/a^3)\epsilon^2/2$ for reasonable values of $b/a$ and $\epsilon$. Exact numerical calculations for the different contributions to the local and global interactions are summarized in Fig.~\ref{fig:pmi3} a) and show good agreement with our analytic approximations. 

As one of the main results of this work Fig.~\ref{fig:pmi3} b) shows the relative magnitudes of local and global spin-spin couplings for different system parameters $\epsilon$ and $b/a$. For this plot we have enhanced the phonon mediated interaction up to the limit $\Delta u/a=0.1$ and have chosen similar signs for direct and phonon mediated contributions. Under these conditions we see that the local coupling can by far exceed residual direct spin-spin interactions.

In summary  we have shown that by employing local modes in our effective spin-spin model the local couplings can be amplified over the residual global interactions in the crystal.
While in principle the ratio $U_0/U_\text{res}\sim 1/\epsilon$ can be made arbitrary large it is in practice limited by the increase of the overall gate time. 
Nevertheless, rations of  $U_0/U_{\rm res}\approx 100$ can already be obtained for reasonable values of $\epsilon\lesssim0.1$ which still allows us to exploit more than $1/10$th of the dipole-dipole interaction $D/a^3$ for gate operations.
The local PMI is enhanced by the local direct coupling $\sim 1/(b/a)^3$ but even for large dipole ratios $\epsilon\approx\Delta\bar{u}$ it exceeds the latter, as can be seen from (\ref{4:U0}), and shows the advantage of our model over purely direct dipole-dipole interaction.

 \subsection{Implemention using optically trapped molecules}
While so far our discussion of the gate operation has been quite general we finally want to address several issues specific for an implementation of these ideas using optically trapped molecules.  We start by giving an estimate of the influence of the tweezer potential which is used to guide the marker molecules.
Since the marker and the crystal are made up of a similar type of molecules and spacings in our model are typically in the order of optical wavelengths the tweezer potential is also seen by molecules in the register.
In a good approximation the potential of a tweezer focused at position $(x_0,y_0)$ is given by
\begin{align}\label{Vtw}
V_\text{tw}(x,y)=V_0 \exp\left[-\frac{(x-x_0)^2+(y-y_0)^2}{2\sigma_\text{tw}^2}\right],
\end{align}
where $V_0=m \omega_\text{tw}^2\sigma_\text{tw}^2$ is the potential depth, $\sigma_\text{tw}$ the width, and $\omega_\text{tw}$ the resulting trapping frequency.
To realize a controlled transition from a regime where the marker is only trapped by the tweezer to one where it is locked to a register molecule (see Fig.~\ref{protocol}), we require that the width of the marker wave function in the tweezer potential is below the crystal lattice spacing, $(\hbar/2m\omega_\text{tw})^{1/2}\leq a$.
We expect little influence on the MDC if the dipole-dipole repulsion in the crystal exceeds the maximal force exerted on register molecules by the tweezer, $3\sqrt{e} D/a^4\gg m\omega_\text{tw}^2\sigma_\text{tw}$.
These two conditions define an allowed region for $\omega_\text{tw}$ which gives the timescale for our adiabatic manipulations.

For  the example of LiCs molecules in an optical trap,  induced dipole moments of about 3 Debye and a value of $r_d=30$ correspond to crystal lattice spacings of $\sim 630\,\text{nm}$. For perpendicular optical trapping frequencies $\omega_\perp = 150\, \text{kHz}$ such lattice spacings allow for minimal values of $b/a\sim 0.63$ which can still be achieved with convenient laser wavelengths $\sim 600$ nm. In such a setup the strength of the dipole-dipole interaction in the crystal is $v_{dd}^{(0)}\sim 35\,\text{kHz}$ such that reasonable ratios of $U_0/U_\text{res} > 100$ give absolute values for the local interaction of $U_0\sim 10\, \text{kHz}$.  The required deep optical trapping potentials can be realized with detunings of the order of $100\,\text{GHz}$ and Rabi frequencies of the order of $100\, \text{MHz}$. For red-detuned trapping fields this will lead to spontaneous photon scattering rates of about $\Gamma_\text{eff}\sim 1\,\text{Hz}$ which could be further reduced using blue detuned lasers. At the same time a tweezer that is focused to $\sigma_\text{tw}\gtrsim 1\mu$m still allows for trapping frequencies $\omega_{\rm tw}$ between approximately $0.6\, \text{kHz}$ and $> 6\,\text{kHz}$.  This is compatible with the general time-scale of the effective spin-spin interaction and therefore adiabatic manipulations of the tweezer  potential do not impose additional limitations on the speed of gate operations.

We point out that the timescales in our model can be greatly enhanced for stronger perpendicular trapping frequencies. For example, by using electrostatic traps in on-chip setups trapping frequencies exceeding 1 $\text{MHz}$ can in principle be achieved ~\cite{AndreNatPhys2006}. For such strong perpendicular confinement intermolecular spacings below $300\,\text{nm}$ could be realized with reasonable ratios of $b/a\sim 0.5$, allowing for values of the local interaction strength easily up to $100\,\text{kHz}$ while keeping $U_0/U_\text{res}\geq 100$.

We finally comment on the readout of the quantum register. In contrast to atoms or ions the lack of closed optical cycling transitions in general prevents a non-destructive detection of molecular qubits. In previous proposals a state selective ionization of the molecules followed by detection of the charged molecule has been proposed as a readout scheme~\cite{QIPDeMilleOriginal}. In our setting a similar method can be applied where in addition right before the readout stage a short time of free expansion of the crystal can be used to separate the molecules. Recently, a class of molecules with almost closed cycling transitions has been identified and used for laser cooling~\cite{ShumanNature2010}. This would then also provide a non-destructive readout by first mapping the state of a register molecule onto such a `closed-cycling' molecules using the gate schemes discussed above.  

\section{Conclusions $\&$ outlook}\label{sec:5}
In summary we have proposed and analyzed a new approach for implementing quantum gate operations between molecular qubits embedded in a MDC. We have shown how arbitrary state dependent and time varying dipole moments for molecular spin qubits can be induced using microwave fields and how this can be exploited for the realization of enhanced phonon-mediated two qubit operations. Compared to previous schemes our proposal does not rely on a spectral or spatial resolution of individual qubits and single site addressability is instead achieved with the help of marker molecules to modify the local phonon structure of the crystal. Therefore, this technique is in particular suited for molecules prepared in self-assembled MDCs, although it can be equally well applied to optical lattices or microtrap arrays.

The general two qubit operation between marker and register molecules which we have analyzed here in more detail can serve as a basic building block for various scalable quantum computing schemes with polar molecules. For an improved performance, different generalization can be envisioned, for example, the use of whole arrays of marker molecules to perform many  operations in parallel  or the multi-layer configurations to change from a 2D to a 3D architecture. Along these lines, also applications for digital quantum simulations \cite{IonQSimu, QStheo} could be considered, where the time evolution of arbitrary spin models is simulated by a stroboscopic application of two qubit gate operations.  Here the  flexibility of moving marker molecules over several lattice sites between the operations can be used to adjust the effective range of the resulting spin-spin interactions.

\section{Acknowledgments} The authors are grateful for help received from Viktor Steixner and Oleg Gittsovich. This work was supported by AFOSR, MURI, the Austrian
Science Foundation (FWF) through SFB F40 FOQUS and by the EU Networks NAMEQUAM and AQUTE.
Y. L. Zhou acknowledges support by Hunan Provincial Innovation Foundation For Postgraduate and NSFC grant No. 11074307.


\end{document}